\documentclass[journal]{IEEEtran}

\usepackage{amsmath,amssymb,amsfonts,amsthm}
\usepackage{xcolor}
\usepackage{cite}
\usepackage{graphicx} 
\usepackage{algorithmic,algorithm}
\usepackage{extarrows}
\usepackage{comment}
\usepackage{subfig}
\usepackage{float}

\ifCLASSINFOpdf
\else
\fi

\begin{document}
\title{A Physical Layer Analysis of Entropy in Delay-Based PUFs Implemented on FPGAs}

\author{Jim~Plusquellic,~\IEEEmembership{Member,~IEEE}
        Jennifer~Howard,~\IEEEmembership{}%
        Ross~MacKinnon,~\IEEEmembership{}%
        Kristianna~Hoffman,~\IEEEmembership{}%
        Eirini~Eleni~Tsiropoulou, ~\IEEEmembership{Associate Member,~IEEE}, 
        Calvin Chan% <-this % stops a space
\thanks{Manuscript received Jan 17, 2022; revised xxx, 20xx.}}

\markboth{Journal of \LaTeX\ Class Files,~Vol.~14, No.~8, August~2015}%
{Shell \MakeLowercase{\textit{et al.}}: Bare Demo of IEEEtran.cls for IEEE Journals}
\maketitle

\begin{abstract}
Physical Unclonable Functions (PUFs) leverage signal variations that occur within the device as a source of entropy. On-chip instrumentation is utilized by some PUF architectures to measure and digitize these variations, which are then processed into bitstrings and secret keys for use in security functions such as authentication and encryption. In many cases, the variations in the measured signals are introduced by a sequence of components in the circuit structure defined by the PUF architecture. In particular, the Hardware-Embedded deLay PUF (HELP) measures delay variations that occur in combinational logic paths on Field Programmable Gate Arrays (FPGAs), which are composed of a set of interconnecting wires (nodes) and look-up tables (LUTs). Previous investigations of variations in these path delays show that it is possible to derive high quality bitstrings, i.e., those which exhibit high levels of uniqueness and randomness across the device population. However, the underlying source and level of variations associated with the constituent components of the paths remain unknown. In this paper, we apply statistical averaging and differencing techniques to derive estimates for the delay variation associated with an FPGA's basic components, namely LUTs and nodes, as a means of fully characterizing the PUF's source of entropy. The analysis is carried out on a set of 50,015 path delay measurements collected from a set of 20 Xilinx Zynq 7020 SoC-class FPGAs, on which 25 identical instances of a functional unit are instantiated, for a total of 500 instances.
\end{abstract}

\begin{IEEEkeywords}
Analysis of Entropy, Physical Unclonable Functions, FPGAs, Within-die delay variation analysis.
\end{IEEEkeywords}

\IEEEpeerreviewmaketitle

\section{Introduction}
\IEEEPARstart{A} Physical Unclonable Function (PUF) is a hardware security primitive that is capable of generating and reproducing bitstrings and keys for use in authentication protocols and encryption algorithms. PUFs are poised to change the current paradigm of using secure non-volatile memories (NVMs) for the storage of device secrets, e.g., encryption keys. PUFs generate bitstrings by applying challenges to the inputs of an integrated circuit (IC) and then measure variations in specific parametric properties of the circuit, e.g., the propagation delay of signals along logic paths. PUF architectures incorporate error correction or error avoidance techniques as a means of reproducing keys without bit-flip errors, which typically requires the use of helper data. PUFs eliminate the need to store the device secrets in NVM and instead store only the challenges and helper data.

An important distinguishing attribute of PUFs over traditional key generation techniques is its source of entropy, which refers to the randomized nature of the parameters measured by the PUF. Random variations in a parameter such as propagation delay are introduced by imperfections in the manufacturing process. Integrated circuit manufacturing uses a technique called photolithography to create the wires and transistors (components) in the device, but the technique has a finite non-zero tolerance making it impossible to create exact copies of the device. Variations in the line widths of wires and transistors, as well as variations which occur within other manufacturing process steps related to doping, diffusion and oxide growth, manifest as small changes in the electrical properties of the components, e.g., transistor drive strength, wire resistance, capacitive load, etc. These variations make each component unique within each device despite the fact that all devices exhibit equivalent logic behavior. Some PUF architectures incorporate specialized hardware components designed to measure these small electrical differences, e.g., changes in the delay of a signal propagating along a path, as a means of deriving a unique bitstring of 0's and 1's for each device.

The source of entropy leveraged by the PUF is complex because the electrical parameters of multiple devices, each affected by multiple sources of physical variation, contribute to the overall variation in the measured parameter. In this paper, we design a set of experiments that isolate the variations that occur in path delays to the smallest possible set of physical components, namely, LUTs and nodes, within an FPGA. Quantifying the amount of delay variation that occurs in these components provides insight into how entropy accumulates along the entire path. This type of experimentally-derived estimation of entropy can be used to model and predict important statistical quality characteristics, namely uniqueness and randomness, of delay PUFs. Our focus is on analyzing data collected from HELP \cite{Aarestad2013}, but the results are applicable to any type of delay PUF.

In this paper, we define a node as a combination of a wire and a switch within a FPGA, and a path segment as a sequence of one or more nodes that connects to the input of a downstream LUT. Path segments and LUTs are connected in series to define logic paths within the circuit structure of a functional unit, e.g., an implementation of the advanced encryption standard (AES) algorithm. Our goal is to estimate the average level of variation associated with the LUTs and nodes within each path segment by using the measured composite variation of the entire path. The data utilized in the analysis is collected from 500 instances of the HELP \cite{Aarestad2013}, which embeds components of AES as its source of entropy. HELP is instantiated on a set of 20 Xilinx Zynq 7020 28 nm SoC-class FPGAs. A netlist processing algorithm is developed that identifies specific pairings of paths from a set of 50,015 measured by each instance of HELP, that enable isolation of variations associated with the constituent components of the paths. The computed difference in the delays of these path pairings are analyzed using statistical methods to estimate the average delay variation of LUTs and nodes. The analysis is performed at the netlist and physical-routing-layer level of abstraction.

%\hfill mds

%\hfill August 26, 2015

\subsection{Contributions}
The contributions of this paper can be summarized as follows:
\begin{enumerate}
\item The analysis of within-die variations in delay at the path component level of granularity, i.e, node and LUT. 
\item The application of statistical differencing techniques designed to isolate the variation of individual path components.
\item The application of statistical averaging techniques over a large population of path pairings and FPGAs to obtain estimates of the average variation and the variance of the variation in the LUTs and nodes of FPGAs, as well as an estimate of noise levels.
\end{enumerate}

The remainder of this paper is organized as follows. Section \ref{Section:Related_Work} describes related work while section \ref{Section:ExperimentalDesign} provides an overview of the experimental design. Section \ref{Section:PathDelayVariation} presents a series of statistical analyses that incrementally evaluate path component variations, culminating in a statistical characterization of within-die variations in the LUTs and nodes of the entire population of FPGAs used in the analysis. Section \ref{Section:Conclusions} presents a summary and conclusions. 

\section{Related Work} \label{Section:Related_Work}

Most techniques proposed for measuring the impact of process variations on delay leverage a ring oscillator (RO). For example, the authors in \cite{YuFPGA2010} propose a differential technique which utilizes ROs placed in two different locations for characterizing LUT and node delays. The goal of their work is to estimate path delays for variation-aware design methodologies where high performance components are placed at boot-time in regions of the FPGA that can support higher frequency operation. Because of structural constraints, their technique is not able to provide estimates for the delay of a node.  

Launch-capture techniques provide an alternative to RO methods for performing delay characterization, and provide the advantage of including variation introduced by clock skew, which represents an actual component of variation during functional operation. The authors of \cite{WongReconfig2009} propose a technique to measure path delays at high resolutions using the reconfiguration capability of a Xilinx DCM to specify the clock input frequency to an Altera FPGA and then monitor for capture failures on the test paths. Although path delays of various functional units are measured and characterized, including inverter chains, full adders and multipliers, no attempt is made to characterize variation in individual LUTs and nodes. Instead, the goal is to provide information for a variation-aware placement strategy as discussed above.

Timing extraction is another technique proposed for deriving FPGA component delays \cite{mehta2012limit}, \cite{gojman2014grokint}, \cite{gojman2014groklab}, again for the purpose of variation-aware placement. Similar to \cite{WongReconfig2009}, the proposed technique utilizes a frequency-based launch-capture method on the FPGA to measure path delays, which are then used as input to a system of equations with separate variables modeling the component delays of the path. The solutions to the equations yields component delays of the constituent elements including LABs and interconnects. 

Previous work on characterizing variations in FPGAs for applications to PUFs include the work published by \cite{Schaumont2010}, where the authors analyze the impact of circuit-level variation on the frequency behavior of ROs using a set of 125 FPGAs. Within-die variation is analyzed using a five-stage RO which is too coarse to reveal variation that occurs at the LUT or node level of granularity. In \cite{Kaps:2013}, the authors propose a ring-oscillator based PUF that leverages internal delay variations in FPGA LUTs as a source of entropy. The proposed design ensures that routing variations are minimized within the structure of the RO. However, an analysis of per-LUT variation is not reported. 

The analysis carried out in this paper differs from previous work in several ways. First, our overall goal is to derive estimates for within-die variations in FPGA component delays for PUF applications. Chip-to-chip variations are explicitly accounted for and eliminated because including them is detrimental to the statistical quality of the bitstrings. Second, we utilize a compensated path differencing operation over a statistically significant sample (500) of devices to deduce component delays in the netlist structure of an AES implementation, in contrast to a set of simple buffer or inverter paths. Third, the source of entropy is tested in a manner consistent with its functional operation, with multiple path sensitization along paths which fan-out, traverse LUTs programmed with complex logic functions and reconverge at primary outputs. Fourth, the analysis of entropy for PUF applications is fundamentally different than the goals of variation-aware placement, with the former focusing on the identification of features that maximize within-die variations across the device population while the later is focused on creating a complete variation profile for each device. We note that our previous work on entropy analysis presented in \cite{CheEntropy:2017} is focused on a statistical characterization of the generated bitstrings and is distinct from the component-oriented physical-layer entropy analysis presented herein.

\section{Experimental Design}\label{Section:ExperimentalDesign}

\subsection{Functional Unit}\label{SubSection:FunctionalUnit}

\begin{figure*}[t]
    \centering
     \includegraphics[width=5.9in,keepaspectratio=true]{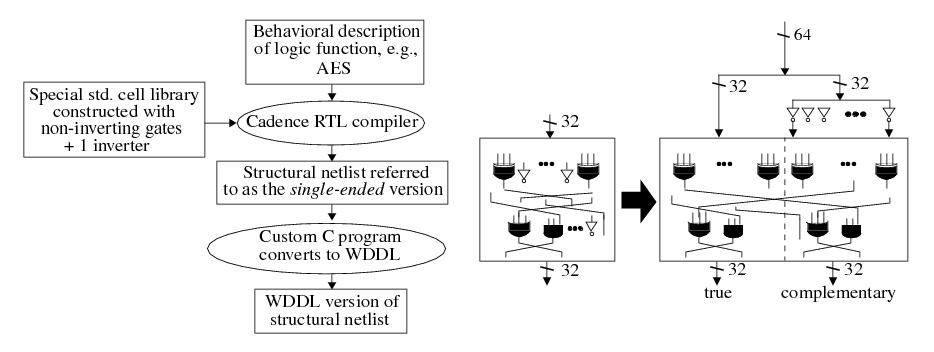}
    \vspace{0pt}
     \caption{Process used to convert behavioral VHDL description of a functional unit to WDDL logic style.}
     \label{Figure:WDDLConversionProcess}
     \vspace{-10pt}
\end{figure*}

The data analyzed in this paper is collected using HELP \cite{Aarestad2013}, which embeds a hazard-free implementation of a combinational logic function called \textit{aes\_mixedcol} as its source of entropy. Here, hazards are defined as additional transitions that occur naturally in combinational logic from differences in the delay of signals along reconverging signal paths. Hazard-free refers to circuit design styles that prevent these transitions from occurring. A VHDL description of the \textit{aes\_mixcol} functional unit, which includes one 32-bit column of the Advanced Encryption Standard (AES) algorithm, is processed through a suite of CAD tools and custom C code to a netlist (schematic) representation. The CAD tool flow is shown on the left side of Fig. \ref{Figure:WDDLConversionProcess}. The behavioral description of the 32-bit column of AES is used as input to the Cadence Register Transfer Level (RTL) compiler along with a special standard cell library discussed below. The RTL compiler produces a netlist from the behavioral code that is referred to as the \textit{single\_ended} version. A custom C program is then used to process the \textit{single\_ended} version into an alternative hazard-free representation of the netlist referred to as the \textit{WDDL} version. WDDL or wave differential dynamic logic was proposed in \cite{Tiri:2004} as a differential power analysis resistant logic style. The WDDL representation guarantees hazard-free signal transitions on the outputs of the \textit{aes\_mixedcol} combinational logic component, which makes it possible to obtain accurate and reproducible measurements of path delays in the hardware experiments. The WDDL representation also allows entire paths to be tested with one transition type, either rising or falling, which will be leveraged later to increase the accuracy of the proposed statistical processing techniques.

\begin{figure}[t]
    \centering
     \includegraphics[width=3in,keepaspectratio=true]{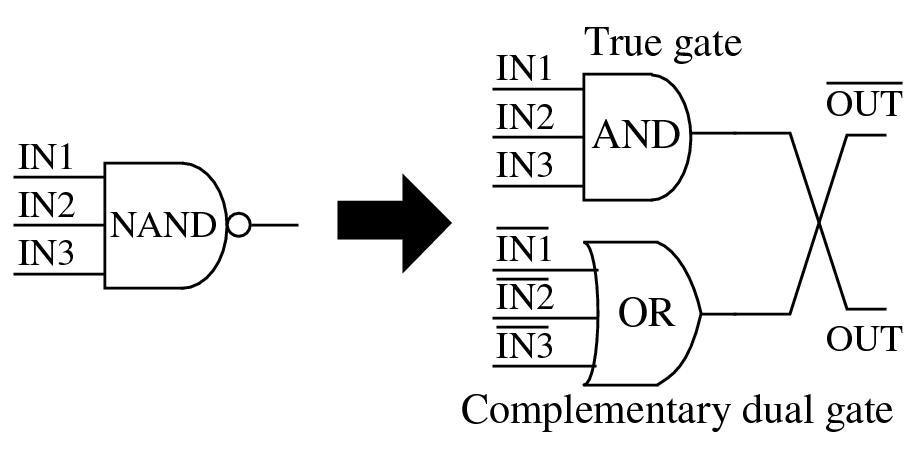}
             \vspace{0pt}
     \caption{WDDL conversion process for an inverting gate.}
     \label{Figure:WDDLGateConversion}
     \vspace{-10pt}
\end{figure}

A functional unit implemented with WDDL logic has both a true and complementary network, as shown on the right side of Fig. \ref{Figure:WDDLConversionProcess}, and possesses twice the number of primary inputs and primary outputs as the \textit{single\_ended} version of the design. WDDL is hazard-free because the two networks are constructed using only non-inverting logic gates. The inversion requirements of logic functions implemented in the functional unit are satisfied in WDDL by cross-connecting between the true and complementary network structures. The special standard cell library used in behavioral synthesis is composed of non-inverting 2-input through 6-input logic gates, e.g., ANDs, ORs. A custom C program is used to convert the single-ended netlist into a WDDL version, which is accomplished by adding the complementary network while simultaneously eliminating the inverters. The WDDL gate conversion process is illustrated in Fig. \ref{Figure:WDDLGateConversion}, where a 3-input NAND gate is translated into an AND gate plus its dual (complementary) OR gate with inputs inverted and outputs swapped as a means of implementing the inversion in the original netlist.

\begin{figure*}[t]
    \centering
     \includegraphics[width=5.8in,keepaspectratio=true]{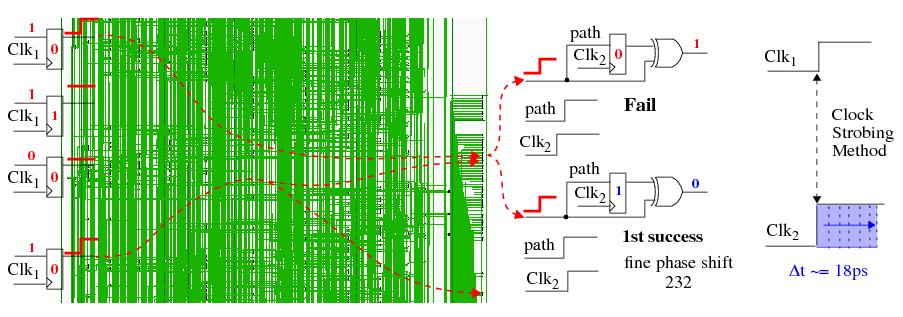}
             \vspace{0pt}
     \caption{Partial Vivado schematic of experimental design illustrating Clock Strobing method used to make high resolution measurements of path delays.}
     \label{Figure:AES_MIXEDCOLSchematic}
     \vspace{-10pt}
\end{figure*}

The WDDL netlist is then included as a component in a behavioral wrapper as shown on the left side of Fig. \ref{Figure:AES_MIXEDCOLSchematic}. The portion shown in green is a snapshot of the elaborated design (schematic) generated by the Xilinx Vivado CAD tool. The wrapper adds components that enable accurate measurements of path delays through the netlist. In particular, the wrapper includes a row of Launch FFs which connect to the primary inputs of the WDDL logic circuit and a row of Capture FFs, which are connected to its primary outputs. The Launch FFs are driven by the primary system clock, $Clk_1$, while the Capture FFs are driven by a special dynamic fine-phase-shifted clock, $Clk_2$.

 WDDL is a two phase logic style with precharge and evaluation phases. During precharge, zeros are applied to all primary inputs which causes a wave of zeros to propagate through the combinational logic of the functional unit. During evaluation, a set of true and complementary input values are applied to the primary inputs which propagate and generate true and complementary primary output values. When operated under these conditions, side-channel leakage is minimized. However, the testing requirements to maintain only hazard-free operation can be relaxed over that proposed by \cite{Tiri:2004}, as exemplified by the input vector assignment given on the left side of Fig. \ref{Figure:AES_MIXEDCOLSchematic}. Here, arbitrary steady-state assignments of '0' or '1' can be assigned to the steady-state inputs, as opposed to only zero. This greatly expands the set of 2-vector sequences that can be applied to the FF inputs while maintaining hazard-free operation. Note that the second WDDL constraint must be maintained, i.e., FF inputs that transition in the 2-vector sequence must all transition in the same direction, either all rising or all falling.

\subsection{FPGA Implementation}\label{SubSection:FPGAImplementation}

\begin{figure}[t]
    \centering
     \includegraphics[width=3.0in,keepaspectratio=true]{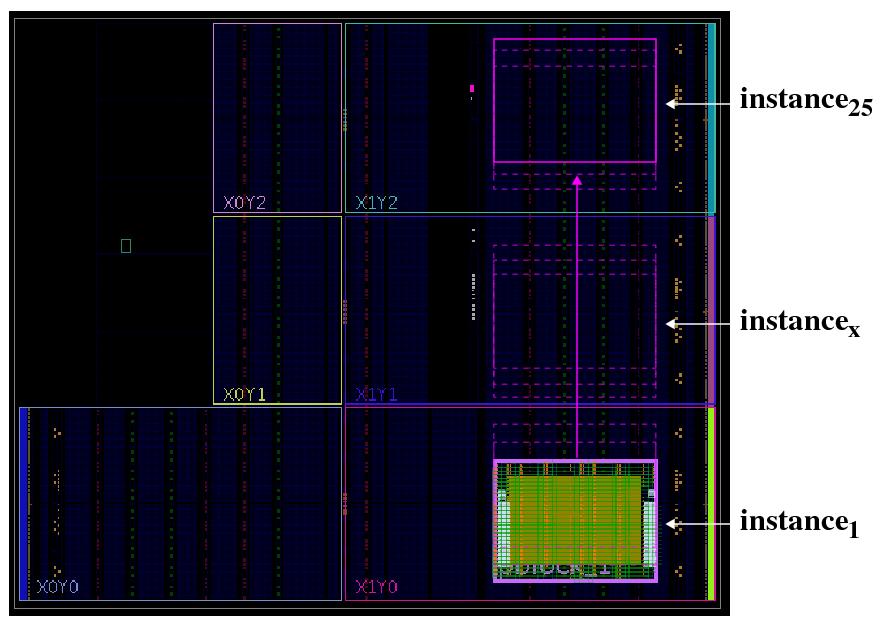}
             \vspace{0pt}
     \caption{Implementation view of \textit{aes\_mixedcol} functional unit (hard macro) and the placement locations where exact copies of the hard macro were placed to create additional instances.}
     \label{Figure:AES_MIXEDCOLLayout}
     \vspace{-10pt}
\end{figure}

Xilinx Vivado is used to implement the design on a set of 20 Zynq 7020 SoC-class FPGAs. The WDDL netlist is first modified by adding a set of wiring constraints to prevent Vivado from applying optimizations. The WDDL netlist and wrapper are then synthesized and implemented into a Vivado \textit{pblock}, which effectively creates a hard macro of the design. The pblock enables the implemented design to be 'locked down' and moved to other regions of the programmable logic fabric while preserving the exact same placement and routing of the original implementation. The original position of the implemented design is labeled $instance_1$ in Fig. \ref{Figure:AES_MIXEDCOLLayout}, with the magenta rectangle representing the pblock. In a series of subsequent synthesis operations, we create 24 additional versions of the pblock, each offset to a unique $y$ coordinate as shown by series of overlapping magenta rectangles labeled $instance_x$ through $instance_{25}$. The pblocks are saved as \textit{design check points} or \textit{DCPs} and each is embedded in a separate parent design that incorporates the timing engine (discussed below). Therefore, we generated as set of 25 programming bitstreams, each used in a sequence of programming operations and experiments on a set of 20 copies of the Zynq 7020 FPGAs. A set of 50,015 path delays were measured for each of these 500 PUF instances.  

\subsection{Measuring Path Delays}\label{SubSection:MeasuringPathDelays}

Automatic test pattern generation (ATPG) is used to derive a set of 1,403 input vector pairs that test 25,015 unique paths with rising transitions and 25,000 unique paths with falling transitions, for a total of 50,015 path delay tests. The vector pairs are applied, one-at-a-time, to the primary inputs using the Launch FFs by applying the first vector of each pair, allowing time for the primary outputs to settle, and then applying the second vector for the timing operation. The input vector sequence introduces transitions into the combinational logic that propagate along a set of paths to a subset of the primary outputs. The Capture FFs shown on the right side of Fig. \ref{Figure:AES_MIXEDCOLSchematic} are used to determine the propagation delay of these paths. The Capture FFs are driven by a second clock output of the Xilinx MMCM (digital clock manager), $Clk_2$, that can be phase shifted by small $\Delta$-t increments of approximately 18 ps at run time. The system clock output, $Clk_1$, is held at 0 phase, as shown by the timing diagram in Fig. \ref{Figure:AES_MIXEDCOLSchematic}. 

The process of timing a path is illustrated in Fig. \ref{Figure:AES_MIXEDCOLSchematic} for one of the primary outputs that experience a transition under the input vector sequence. The phase shift of $Clk_2$ is initially set to 0 and is incremented by one $\Delta$-t increment as the input vector sequence is applied repeatedly (until all primary outputs with transitions are timed). The region labeled \textbf{Fail} shows the final state of the Capture FF when the transition along a path fails to reach its input before $Clk_2$ is asserted. An XOR gate is used to decide the pass/fail status. Here, the output value of the path under the first vector remains stored in the Capture FF, and the XOR gate produces a '1' because the output value under the second vector is '1'. As subsequent tests are applied with $Clk_2$ configured with increasingly larger phase shifts, eventually, the transition arrives at the Capture FF before $Clk_2$ is asserted and the Capture FF successfully stores the second vector's output value. The region labeled \textbf{1st success} in Fig. \ref{Figure:AES_MIXEDCOLSchematic} illustrates the final state of the Capture FF under this condition. The digitized delay value for the tested path, e.g., 232, represents the number of $\Delta$-t fine phase shift increments needed to reach this terminal state, i.e., with the XOR gate producing a '0'.

We apply this clock strobing technique using each of the 1,403 input vector sequences. Each vector pair times approximately 35 paths and generates a corresponding timing value. We refer to the digitized timing values as \textbf{path delays} in the following analysis. The clock strobing operation is repeated 16 times for each vector sequence and an average path delay, $PD$, is computed using the 16 samples. The averaging significantly reduces measurement noise as a source of variation (discussed further below). We convert the average path delays from fine-phase-shift increments to seconds by multiplying by 18 picoseconds, and report the delay in picoseconds (ps) or nanoseconds (ns) in the remainder of this paper, e.g., 232.3 * 18 ps = 4.181 ns. All measurements are made at room temperature ($25^oC$) in a temperature chamber and under nominal supply voltage ($1.00V$) conditions.

\subsection{Physical Characterization of Paths and Delays} \label{SubSection:PhysicalCharacterizationPaths}

\begin{figure}[t]
    \centering
     \includegraphics[width=3.4in,keepaspectratio=true]{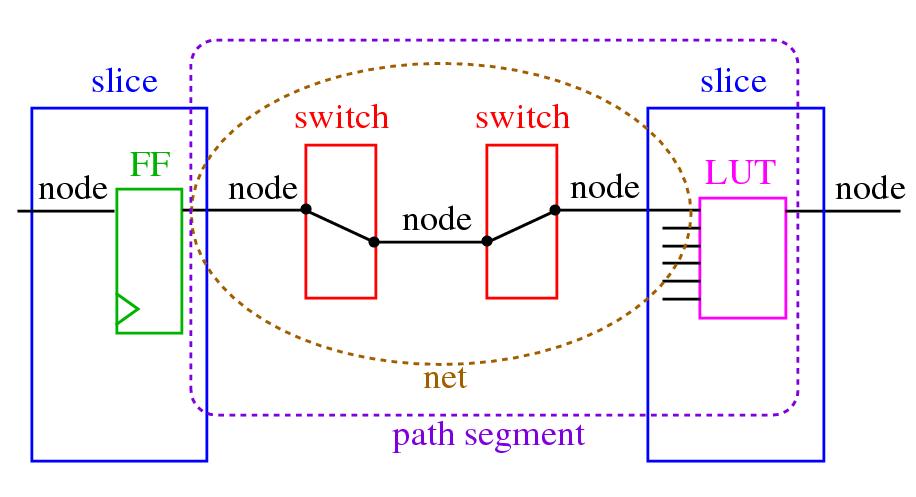}
     \caption{Definitions of FF, node, switch, net and LUT.}
     \label{Figure:DefinitionsFFNodeSwitchNetLUT}
     \vspace{-12pt}
\end{figure}

A path is defined as a sequence of gates connected in series with interconnect wires and switches. Xilinx uses the term look-up table or \textit{LUT} in reference to a gate and \textit{node} to refer to a wire as shown by the illustration in Fig. \ref{Figure:DefinitionsFFNodeSwitchNetLUT}. LUTs and FFs are contained inside of slices. Switches are always interposed between nodes. The term \textit{net} is used in reference to a sequence of nodes-switch combinations and the term \textit{path segment} is used in reference to a net-LUT combination as shown in the figure. The delay of a path is measured between a Launch FF and a Capture FF, and can be partitioned into a set of \textit{path segment} delays by dividing the path delay by the number of LUTs + 1 to account for the Launch FF \textit{clk-to-q} delay. Similarly, a \textit{path segment} delay can be partitioned into a set of node delays by dividing the \textit{path segment} delay by the number of \textit{nodes-switch} sequences. Given that we cannot separate node and switch delays, we refer to the node-switch delay as simply \textit{node delay}, recognizing that it represents the delay of both. Later, we will describe a statistical differencing technique that is able to estimate the node-switch delays. Similarly, we cannot separate LUT delays from FF delays and therefore our estimates of node and LUT delays are likely to be slightly biased given these limitations. However, we validate our estimates of node and LUT delays using several techniques and conclude that they represent good estimates of a path's constituent component delays.

\subsection{Path Node Expansion}\label{SubSection:PathNodeExpansion}

The node names associated with each net name in the netlist definition of the path are obtained by running a custom TCL script in Xilinx Vivado. The TCL script reads each of the 50,015 netlist strings that define the paths, one at a time, and performs a series of calls within Xilinx Vivado to obtain the detailed routing information associated with each net name in the netlist string. As an example, the following lists the first four net names associated with one of the tested paths (netlist string) from the set of 50,015.

\begin{enumerate}
\item Row1\_in/GEN\_SFF[17].MUXDScanFFEle/ScanEle 
\item i\_2203\_48 
\item n\_71\_inferred\_i\_1 
\item n\_215\_inferred\_i\_1 
\item (remaining net names for the first path)...
\end{enumerate}

The TCL script expands each net name into a set of node names, given as follows for the first net name from the path defined above.

\begin{enumerate}
\item CLBLM\_L\_X52Y19/CLBLM\_L\_BQ 
\item CLBLM\_L\_X52Y19/CLBLM\_LOGIC\_OUTS1 
\item INT\_L\_X52Y19/NN2BEG1 
\item INT\_L\_X52Y21/SR1BEG1 
\item INT\_L\_X52Y20/SL1BEG1 
\item INT\_L\_X52Y19/IMUX\_L42 
\item CLBLM\_L\_X52Y19/CLBLM\_L\_D6
\end{enumerate}
%\textcolor{green}{How do you know which things in the list above are LUTs vs nodes?} Jen: The list above are node names. The first list are net names. The LUTs are not present in the netlist string, only the sequence of net names. Every net name is followed by a LUT, except the last, which is followed by a Capture FF. Hopefully, I've cleared this up above with the definitions (it was definitely confusing -- sorry about that) Jim: I think I get it now. This description and the new paragraph help, thanks  To make sure I get it, the path described in the lists would be Row1\_in/GEN\_SFF[17].MUXDScanFFEle/ScanEle -> LUT -> i\_2203\_48 -> LUT ....., right?  For the second list, is there a switch in between each node?

\begin{figure}[t]
    \centering
     \includegraphics[width=3.0in,keepaspectratio=true]{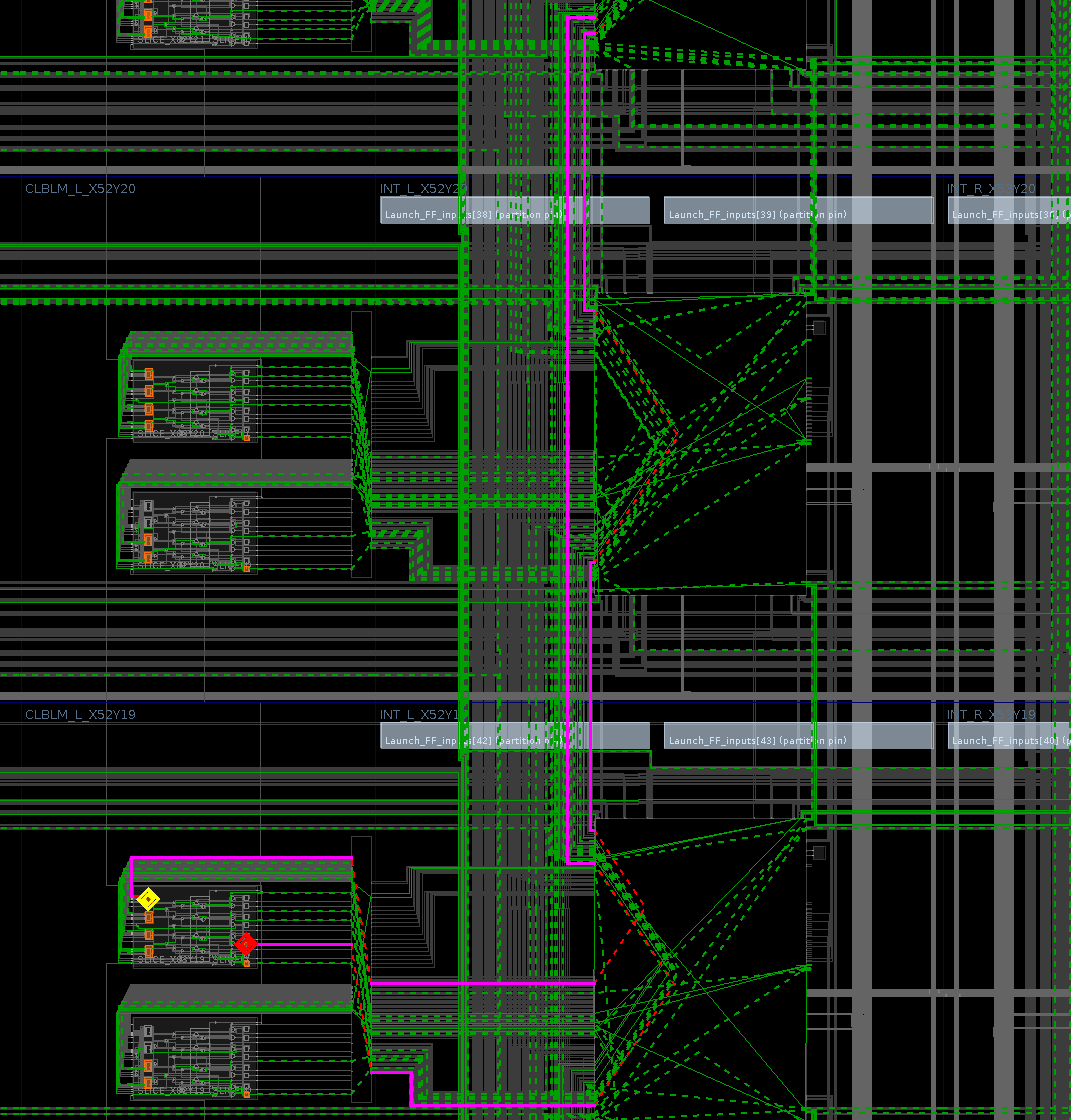}
     \caption{Highlighted Launch FF (red diamond), nodes (magenta) and terminating LUT (yellow) associated with a path segment.}
     \label{Figure:FirstPathFirstNet}
     \vspace{-10pt}
\end{figure}

Each net name is expanded in a similar fashion. Fig. \ref{Figure:FirstPathFirstNet} shows the Xilinx Vivado layout for this example path segment. The path begins with a Launch FF shown as a red diamond. The \textit{path segment} follows and consists of a series of nodes shown as solid magenta wires and red dotted lines through switches, and ends with a terminating LUT (yellow diamond). The node names given above refer to the solid magenta wires in the figure.

\section{Analysis of Path Delay Variation}\label{Section:PathDelayVariation}

\subsection{Sources of Variation and Compensation}\label{SubSection:VariationAndCompensation}

The goal of our analysis is to estimate the average within-die variation introduced by the LUTs and nodes using measurements of delay for the entire path. The path delays for a device are affected by three primary sources of variations; measurement noise, global (chip-to-chip) process variations and within-die process variations. Measurement noise can be significantly reduced by measuring each path delay multiple times and computing an average. In our analysis, we compute an average path delay for each of the 50,015 paths from a set of 16 samples collected for each path. 

Although global process variations represent a source of Entropy for PUFs, they cannot be used in large scale deployment of the PUF because global variations impact all path delays in a similar fashion. Therefore, in large chip populations the probability that two chips experience similar global process variations becomes large, thereby resulting in chips with similar corresponding path delays. This, in turn, will significantly reduce the uniqueness characteristics of the bitstrings in cases where chip-to-chip variations are leveraged by the PUF's bitstring generation algorithm.  
% Jim: Also mention LARGE correlations in the response bitstrings here.

\begin{figure*}[t]
    \centering
     \includegraphics[width=4.0in,keepaspectratio=true]{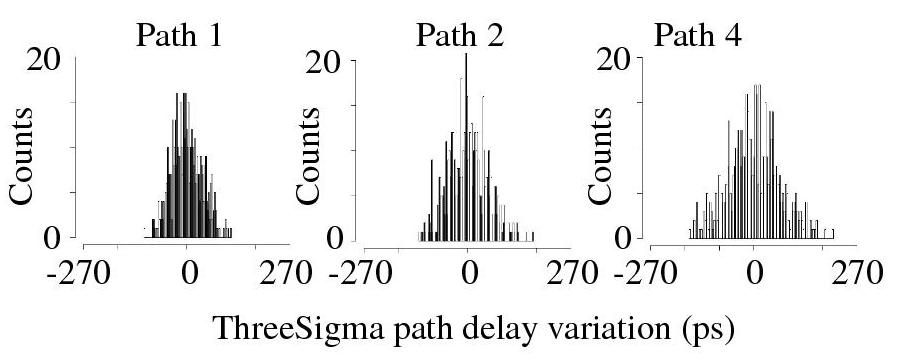}
     \caption{Distributions of path delay variations for three paths across 500 PUF instances.}
     \label{Figure:P1_P2_P4_PathVariationDist}
     \vspace{-12pt}
\end{figure*}

Therefore, the bitstring generation algorithm used by the HELP PUF removes chip-to-chip process variations and utilizes only within-die variations. Within-die path delay variations are what remains after measurement noise and global delay variations are removed. We remove global delay variations by applying a pair of linear transformations to the sample-averaged path delays measured from each of the 500 PUF instances. First, the mean and standard deviation of the measured path delay distributions are computed for each PUF instance using Eqs. \ref{ChipMean} and \ref{ChipStdDev}. Here, $PD_{i, j}$ represents the average path delay (each computed using 16 samples as described above) for PUF instance $i$, path $j$ and $NP$ represents the number of paths (50,015 in our experiments).   

\begin{equation}
    \mathbf{u}_i = \frac{\sum\limits_{j=1}^{NP}PD_{i,j}}{NP}
    \label{ChipMean}
\end{equation}

\begin{equation}
    \mathbf{\sigma}_i = \sqrt\frac{\sum\limits_{j=1}^{NP}(PD_{i,j} - \mathbf{u_i})^2}{NP - 1}
    \label{ChipStdDev}
\end{equation}

The mean and standard deviation for each PUF instance are then used to standardize each of the values in the PUF instance distribution by subtracting the mean and dividing by the standard deviation, i.e., the path delays are converted to standard normal form, using Eq. \ref{PDZ}. 

\begin{equation}
    Z_{i,j} = \frac{(PD_{i,j} - \mathbf{u}_i)}{\mathbf{\sigma}_i}
    \label{PDZ}
\end{equation}

The next step is to calculate the mean of the PUF instance means and the mean of the PUF instance standard deviations using Eqs. \ref{ChipMeanofMeans} and \ref{ChipMeanofStdDev}. Here, NC represents the number of PUF instances (500 in our experiments).

\begin{equation}
    \mathbf{u}_{ref} = \frac{\sum\limits_{i=1}^{NC}\mathbf{u}_i}{NC}
    \label{ChipMeanofMeans}
\end{equation}

\begin{equation}
    \mathbf{\sigma}_{ref} = \frac{\sum\limits_{i=1}^{NC}\mathbf{\sigma}_i}{NC}
    \label{ChipMeanofStdDev}
\end{equation}

The second linear transformation multiplies each of the standardized path delays by the mean standard deviation and then adds in the mean of the means, using Eq. \ref{PDC}.  

\begin{equation}
    \mathbf{PDC_{i,j}} = Z_{i,j}*\sigma_{ref} + u_{ref}
    \label{PDC}
\end{equation}

The effect of these transformations is to skew and scale all PUF instance distributions such that the overlap between them is maximized and nearly all of the global, chip-to-chip delay differences are removed. In other words, all PUF instance distributions are scaled to the mean performance delay of the population. This process enables the analysis of within-die delay variations for each of the paths across the PUF instance population. We use the term within-die delay variation for a path despite the fact that the analysis is carried out on the set of delays measured across the PUF instances. 
%Note that within-die delay variation defined in this manner actually represents entropy for the PUF. 
The application of the linear transformations is called compensation, and we use the term $PDC$ to refer to these compensated path delays in the following sections.

\subsection{Single Path Variation Analysis}\label{SubSection:SinglePathVariationAnalysis}

As a basis for the analysis that follows, we first characterize delay variation for each of the 50,015 individual paths. For each path, the mean, standard deviation and variance of the delays measured across all 500 PUF instances are computed using Eqs. \ref{SinglePathMean}, \ref{SinglePathStdDev} and \ref{SinglePathVariance}, respectively. In the equations, $i$ refers to PUF instance, $j$ refers to path, and $NC$ refers to the total number of PUF instances. Fig. \ref{Figure:P1_P2_P4_PathVariationDist} shows a histogram of the compensated path delay variations, $K_{i,j}$, across the 500 PUF instances for three paths from the set of 50,015 paths, each computed using Eq. \ref{SinglePathDelayVariation}. The Entropy associated with these distributions is best expressed as 3 * $\mathbf{\sigma_{PDC_j}}$, i.e., the ThreeSigma standard deviation, as given by Eq. \ref{SinglePathThreeSigma}. In particular, the ThreeSigma standard deviation computed for the path distributions in Fig. \ref{Figure:P1_P2_P4_PathVariationDist} are 115 ps, 154 ps and 196 ps, respectively.

\begin{equation}
    \mathbf{u_{PDC_j}} = \frac{\sum\limits_{i=1}^{NC}PDC_{i,j}}{NC}
    \label{SinglePathMean}
\end{equation}

\begin{equation}
    \mathbf{\sigma_{PDC_j}} = \sqrt\frac{\sum\limits_{i=1}^{NC}(PDC_{i,j} - \mathbf{u_{PDC_j}})^2}{NC - 1}
    \label{SinglePathStdDev}
\end{equation}

\begin{equation}
    \mathbf{\sigma^2_{PDC_j}} = \mathbf{(\sigma_{PDC_j})^2}
    \label{SinglePathVariance}
\end{equation}

\begin{equation}
    \mathbf{K_{i,j}} = PDC_{i, j} - u_{PDC_j}
    \label{SinglePathDelayVariation}
\end{equation}

\begin{equation}
    \mathbf{ThreeSigma_{PDC_j}} = 3 * \mathbf{\sigma_{PDC_j}}
    \label{SinglePathThreeSigma}
\end{equation}

%\textcolor{green}{How/Why does ThreeSigma directly reflect Entropy?} 

%\textcolor{blue}{Jen: Plus and minus ThreeSigma IS the actual width of the measured variations (unlike variance which is the square of the std. dev.) The HELP bitstring generation algorithm maps these distributions directly to 0's (left half) and 1's (right half) in the bitstrings. }
%\textcolor{green}{Jim: Oh, this was helpful.  This explanation at the end of the report is what I was looking for.  thank you }

Although ThreeSigma is a useful metric because it directly reflects Entropy, variance, $\mathbf{\sigma^2_{PDC_j}}$, is more convenient when mathematical operations, e.g., addition and subtraction, are carried out during decomposition. For example, the total variance for an entire path is the sum of the variances of the individual path segments. Similarly, the total variance that results when the delays of two entire paths are subtracted is again the sum of the individual path segment variances. Therefore, we use variance in the following, and report ThreeSigma only in the context of Entropy.

\begin{figure}[t]
    \centering
     \includegraphics[width=3.4in,keepaspectratio=true]{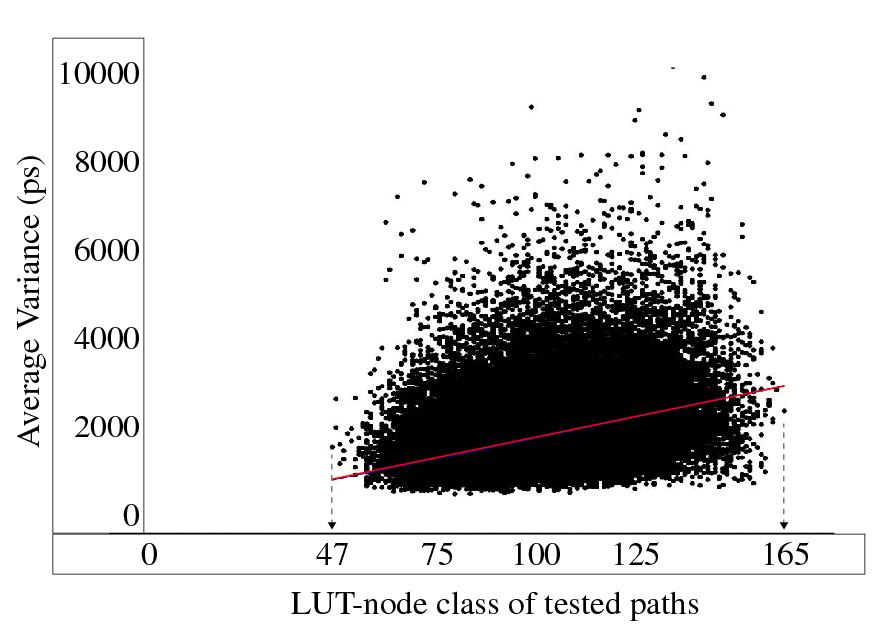}
     \caption{Variance of the 50,015 paths as a function of the number of components in the paths, e.g., nodes + LUTs.}
     \label{Figure:SinglePathSortedVariance}
     \vspace{-12pt}
\end{figure}

The average variance for each of the 50,015 paths is depicted in Fig. \ref{Figure:SinglePathSortedVariance}, sorted along the x-axis according to the number of components (LUTs + nodes) in each of the paths, referred to as the \textbf{LUT-node class}. For example, the shortest path is shown at x-position 47, which consists of 8 LUTs and 39 nodes, while the longest path at 165 is composed of 26 LUTs and 139 nodes. The blob-shape of the distribution indicates that the number of components in the path is not well-correlated to its variation. However, a relationship, albeit weak, does exist that is portrayed by the superimposed red line. The following sections describe the process used to derive this line, which represents the sum of the average variances for each LUT-node class.

%\textcolor{green}{What is a LUT-node class?  Is 47 a LUT-node class?  Does 47 just indicate that there are 47 components that are a mix of LUTs and nodes? Would a path that had 6 LUTs and 41 nodes also be plotted on the same x-axis point? If that is the definition of a LUT-node class, can we add that definition here to help with the following sections?  Then can the x-axis labels on Figs 9 and 10 be changed to LUT-node class?} \textcolor{blue}{Yes to all of these questions}

\begin{figure}[t]
    \centering
     \includegraphics[width=3.4in,keepaspectratio=true]{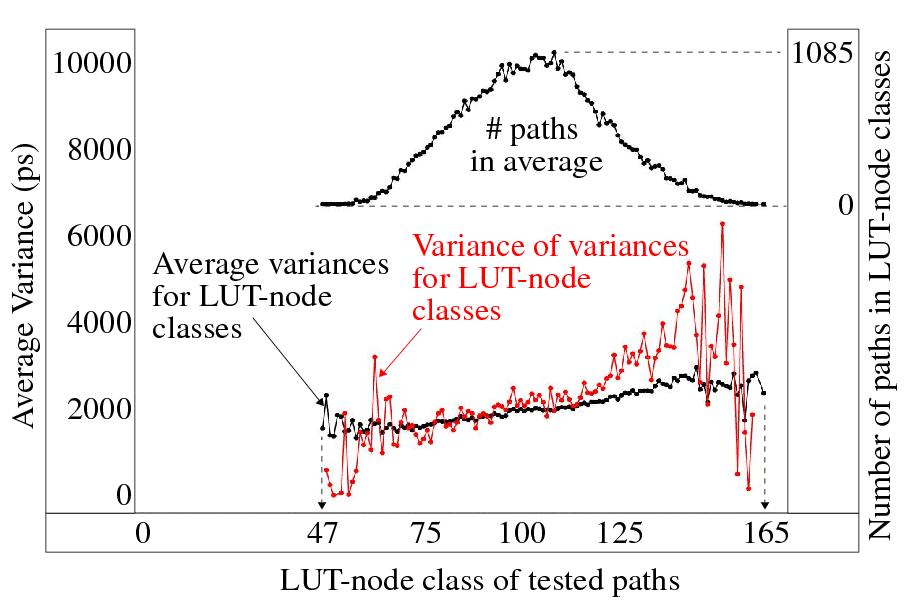}
     \caption{Average variance and 'variance of the variances' for the individual path variances from Fig. \ref{Figure:SinglePathSortedVariance} as a function of the number of components in the paths, e.g., nodes + LUTs.}
     \label{Figure:SinglePathSortedVarianceStats}
     \vspace{-10pt}
\end{figure}

The variances shown in Fig. \ref{Figure:SinglePathSortedVariance} are averaged using the points from each x-axis LUT-node class and plotted as the black curve labeled "Average variances for LUT-node classes" in Fig. \ref{Figure:SinglePathSortedVarianceStats}. Each of these points represents the average variance for paths in each LUT-node class, and is defined using Eq. \ref{MeanSinglePathVariance}. Here $NP_{m}$ represents the number of paths with $m$ components (LUTs + nodes) as given by the curve labeled '\# paths in average', and $\mathbf{\sigma^2_{PDC_{p}}}$ represents the %average
variance from Eq. \ref{SinglePathVariance} of some path $p$ in the set of paths of length $m$. The red curve in Fig. \ref{Figure:SinglePathSortedVarianceStats} plots the variance of the averaged LUT-node class variances, which will later be used to provide bounds on the uncertainty in the estimates of per-LUT and per-node variance (which is the goal of this paper).

%textcolor{green}{Can we change $NP_{len=m}$ to just $NP_{m}$  The len=m keeps throwing me off when I read the equations.}

\begin{equation}
    \mathbf{u}_{\sigma^2_m} = \frac{\sum\limits_{p=1}^{NP_{m}}\mathbf{\sigma^2_{PDC_{p}}}}{NP_{m}}
    \label{MeanSinglePathVariance}
\end{equation}

\begin{figure}[t]
    \centering
     \includegraphics[width=3.4in,keepaspectratio=true]{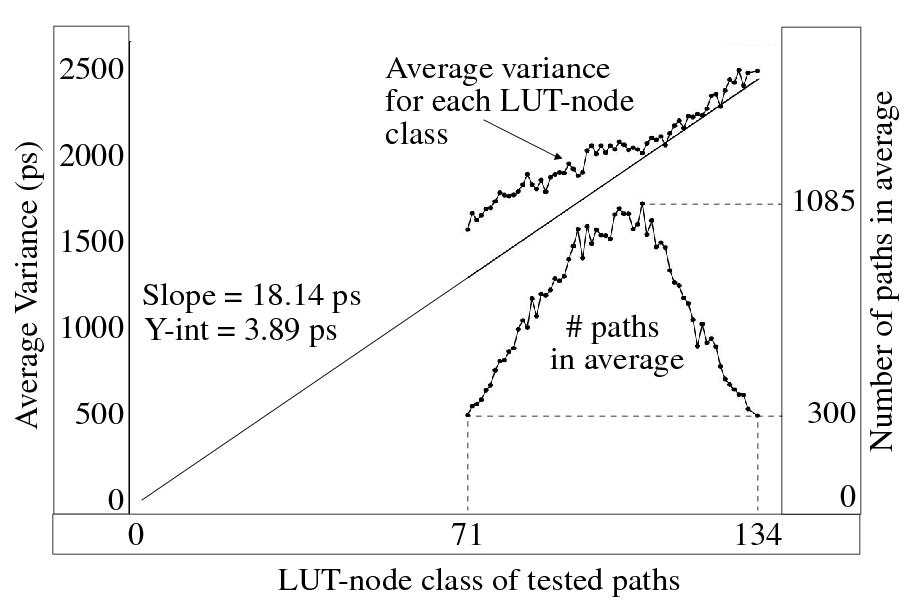}
     \caption{Average variance of the 50,015 paths as a function of the number of components in the paths, i.e., nodes + LUTs. The average variance for x-axis values in which the number of components is less than 300 are excluded. A line representing an estimate of the average variation per LUT and per node is superimposed with scale, slope and intercept shown on the left. The number of paths used in each average is shown by the lower curve with scale on the right.}
     \label{Figure:SinglePathGroupAvePredictionDataBoth}
     \vspace{-10pt}
\end{figure}

Fig. \ref{Figure:SinglePathGroupAvePredictionDataBoth} trims off the edges of the LUT-node class averages shown in Fig. \ref{Figure:SinglePathSortedVarianceStats} and blows up the region of interest. The statistical noise is large for LUT-node classes in which the number of paths in the average fell below 300 (as shown in Fig. \ref{Figure:SinglePathSortedVarianceStats}) and are therefore excluded.

The variances for LUT-node classes below approximately 100 deviate somewhat from the expected linear-through-0 trend portrayed by LUT-node classes larger than 100. This illustrates that the analysis of full length path delays is capable of providing only a rough approximation of node and LUT variances. The line with slope 18.14 is generated using a more accurate estimate of per-LUT and per-node variance derived from an alternative analysis of path pairings presented in the following sections.

\subsection{Path Delay Differences}\label{SubSection:PathDelayDifferences}

The path pairing analysis described here pairs each of the rising path delays with each other, and each of the falling path delays with each other, to create two large groups. The path delays in each pairing are then subtracted. The path pairing analysis enables delay variation to be analyzed at higher levels of resolution. This follows because the paths in each of the pairings can have path components (LUTs and nodes) that are common to both paths. In such cases, the delay difference captures variations in only those components of the two paths that are different, effectively eliminating the delay variation associated with the common components, in the spirit of common mode rejection.

\begin{figure}[t]
    \centering
     \includegraphics[width=3.0in,keepaspectratio=true]{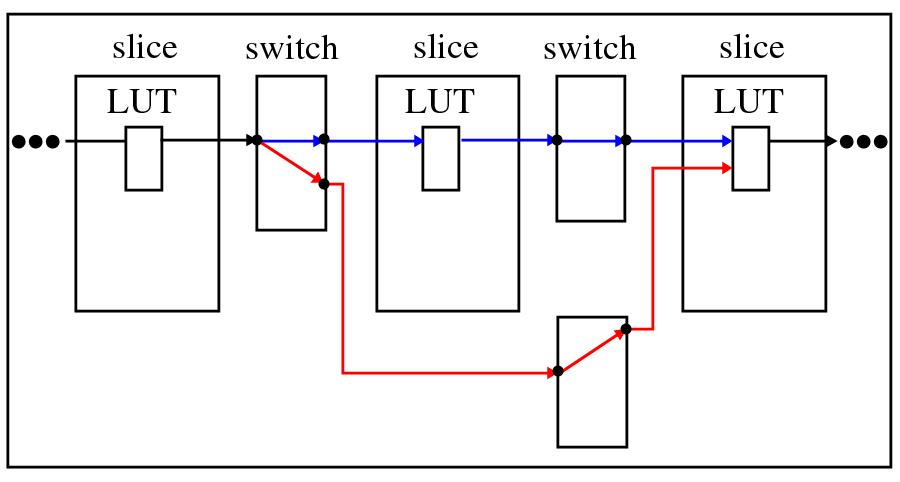}
     \caption{Path pair illustration showing the minimal number of distinct path components, referred to as the OneExtraLUT configuration.}
     \label{Figure:OneExtraLUTTargetPathStructure}
     \vspace{-10pt}
\end{figure}

The illustration in Fig. \ref{Figure:OneExtraLUTTargetPathStructure} shows an extreme example of common mode rejection, which we refer to as the OneExtraLUT path pair configuration. Here, both paths traverse the exact same nodes and LUTs (not shown) through components up to and including the LUT shown on the left in the figure. The two paths fan-out to different nodes in the following switch. The top path highlighted in blue travels through an additional LUT, a second switch, and then reconverges with the bottom path (highlighted in red) at the right-most LUT. The remaining segments of the two paths after the right-most LUT share a common route to the same end point. The difference in the path delays under these conditions capture the variation introduced by the red and blue nodes plus the variation introduced by the additional LUT in the longer (blue) path.  

\begin{figure}[t]
    \centering
     \includegraphics[width=3.0in,keepaspectratio=true]{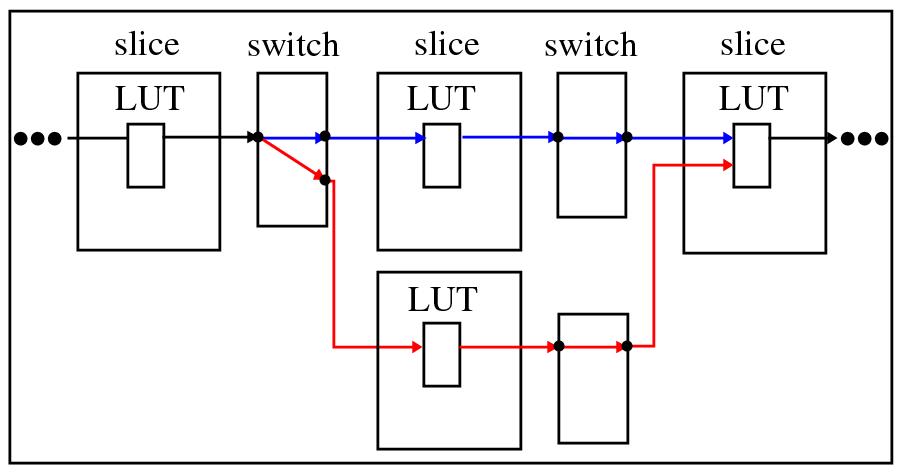}
     \caption{Illustration of the Bubble path pair configuration.}
     \label{Figure:BubbleTargetPathStructure}
     \vspace{-10pt}
\end{figure}

The OneExtraLUT configuration can be extended to include additional distinct components in the path pair. For example, Fig. \ref{Figure:BubbleTargetPathStructure} shows a configuration, called the Bubble configuration, in which the number of distinct LUTs increases to two. In this configuration, the two paths fan-out at the switch on the left, traverse through two different LUTs and reconverge again at the LUT shown on the right. Here, the delay difference variation is expected to be larger than the variation computed for the OneExtraLUT configuration because two LUTs are contributing to the Entropy. 

\begin{figure*}[t]
    \centering
     \includegraphics[width=5.2in,keepaspectratio=true]{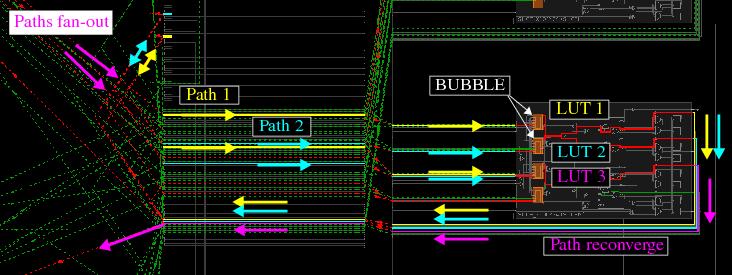}
     \caption{Example of Bubble path pair configuration in the experimental design.}
     \label{Figure:ExampleBubbleTargetPathStructure}
     \vspace{-10pt}
\end{figure*}

One of the Bubble path pairs identified in our experimental design is shown in Fig. \ref{Figure:ExampleBubbleTargetPathStructure}. The two paths fan-out as shown on the left, and then route to the inputs of LUT 1 and LUT 2, respectively, which form the bubble structure. The paths route out of these LUTs, back to the switch and then re-enter the slice to drive the inputs of LUT 3, where the two paths reconverge to a common path highlighted in magenta. In this example there are 12 distinct nodes and 2 distinct LUTs. 

\begin{figure}[t]
    \centering
     \includegraphics[width=3.4in,keepaspectratio=true]{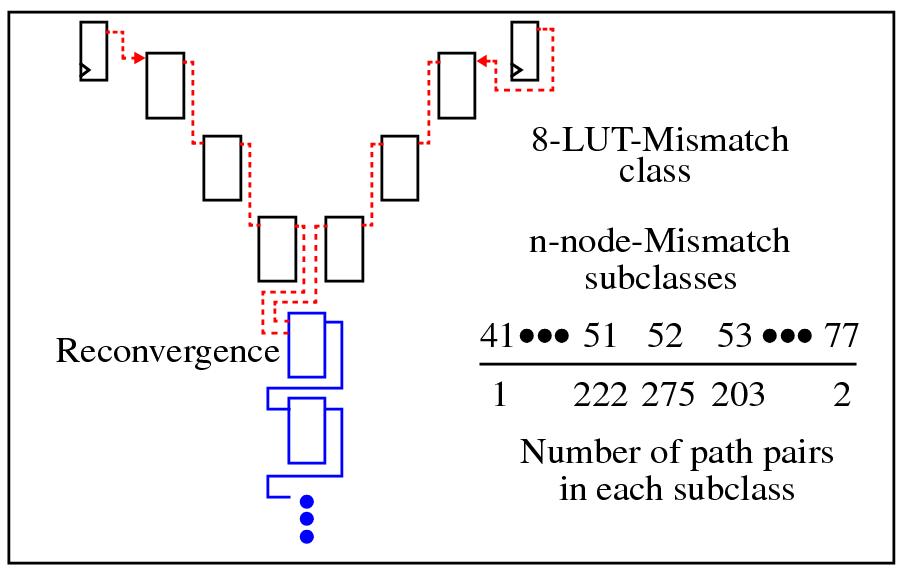}
     \caption{Example of a permissible 8-LUT-Mismatch class configuration with n-Node-Mismatch subclasses.}
     \label{Figure:NMisMatchTargetPathStructure}
     \vspace{-10pt}
\end{figure}

The remaining target path pair configurations are similar to the Bubble configuration except extended to allow additional pairs of LUTs to be distinct in the two paths. We refer to these configurations as \textit{L-LUT-Mismatch} where $L$ is always an even number, e.g., the Bubble configuration is also classified as a \textit{2-LUT-Mismatch}. Although it is possible to allow path pairings with, e.g., 3 distinct LUTs, odd values force the paths to be of different lengths, which acts to add statistical noise particularly when the difference in path lengths becomes large. To address this issue, we constrain our analysis to path pairings in which the two paths are the same length, and additionally require the two paths to share at least one LUT in common, i.e., they always have the same end point capture FF.  An example of a permissible configuration for the \textit{8-LUT-Mismatch} path pairing class is illustrated in Fig. \ref{Figure:NMisMatchTargetPathStructure}. Here, the paths start at different launch points but converge before reaching the end point (Note that the Launch FFs are counted as LUTs in the \textit{L-LUT-Mismatch} class). These restricted configurations help to improve the correlation between variance and path length, which can become poor for arbitrary pairings of paths as shown earlier in reference to Fig. \ref{Figure:SinglePathSortedVariance}.

By extending the number of permissible mismatches in the LUTs defining a path pairing beyond 2, it is possible to find large numbers of instances of these alternative configurations, unlike the OneExtraLUT and Bubble configurations, of which only 20 and 278 instances exist in our design, respectively. The statistical averaging of large numbers of instances acts to average out undesirable effects and enables a better estimate of the delay variance associated with LUTs. Another advantage afforded by the \textit{L-LUT-Mismatch} classes is the ability to isolate and estimate the average delay variance associated with nodes, as discussed in the following.

In order to identify those path pairings that meet the target configurations, we wrote a C program that analyzes all rising and falling path pairing combinations. As discussed in Section \ref{Section:ExperimentalDesign}, the vectors applied to the functional unit tested 25,015 paths with rising transitions and 25,000 paths with falling transitions. Note that the WDDL-style logic used in the functional unit does not allow inversions of signals along a path. So, for example, driving a rising transition on a path input guarantees that all path segments along the path propagate a rising transition. Therefore, we construct path pairings using only paths of the same transition type, i.e. rising with rising or falling with falling, which maximizes the cancellation of variation in those path components that are common to both paths. Eq. \ref{NumRiseFallSeparate} gives the number of path pairings analyzed using this pairing strategy, with $NPR$ and $NPF$ indicating the number of paths with rising transitions and falling transitions, respectively.

\begin{multline}
    \mathbf{|PathPairings|} = (NPR)(NPR-1)/2~+\\
    (NPF)(NPF-1)/2 = 625,350,105
    \label{NumRiseFallSeparate}
\end{multline}

For each of the path pairings in each of the \textit{L-LUT-Mismatch} classes, a node string analysis is carried out. As discussed earlier, each net name is composed of one or more node names, and the number of nodes per net name vary from one path segment to the next. Similar to the analysis presented in the previous section, the individual variances for each \textit{L-LUT-Mismatch} class are partitioned further according to the number of distinct nodes in each path pair, i.e., separate subgroups of path pairs are created within each \textit{L-LUT-Mismatch} class called \textit{N-Node-Mismatch} subclasses. A mean variance is then computed for each of these \textit{N-Node-Mismatch} subclasses. 

The node subgroup partitioning scheme is illustrated for the \textit{L-LUT-Mismatch} class in Fig. \ref{Figure:NMisMatchTargetPathStructure}. The total number of path pairings for this class is 25,316, of which 2,829 are considered permissible configurations. The distinct nodes subject to counting are shown graphically as red dotted lines. A portion of the subgroup partitioning table is shown on the right. For example, subgroup 41 has only one path pair member, while the number of members in subgroups 51, 52 and 53 increase to more than 200. For larger \textit{L-LUT-Mismatch} classes, the number of members can increase to more than 10,000.

The following averaging property is a key component to the process we use to derive estimates for LUT and node variance. In cases where the cardinality in the \textit{N-Node-Mismatch} subclasses is large, the average variance introduced by $L$ distinct LUTs and $N$ distinct nodes is approximately the same because averaging a large number $z$ of random variances with mean $x$ approaches $x$ as $z$ becomes large. Therefore, by pairing subgroups and subtracting their average variances, it is possible to obtain estimates of the average variance of $y$ components, where $y$ represents the difference in $N$ among the two \textit{N-Node-Mismatch} subclasses. The subgroup pairings of greatest interest are those in which the difference in the $N$s of the subclasses is equal to one. In such cases, the variance represented by the difference is that corresponding to only one node, which represents one of the goals of our analysis.

As an example, the difference variance computed by subtracting the average variance of subgroup 52 from 53 in Fig. \ref{Figure:NMisMatchTargetPathStructure} is 0.1106, which represents an estimate of node variance because the average LUT and node variances associated with 52 of the components in these subgroups is approximately the same (by virtue of averaging the variances of 275 and 203 components) and is therefore eliminated in the difference. This process can be repeated for other pairings and the results averaged to obtain an overall estimate of node variance, as we show in the following section. 

\subsection{Statistics of Path Pair Differences}\label{SubSection:StatisticsPathPairDifferences}

A set of statistics are computed using path delay differences, similar to the statistics presented earlier for the individual path delays from Section \ref{SubSection:SinglePathVariationAnalysis}. The path delay compensated differences, $PDCD$, are computed for each PUF instance $i$ and path pairing ($j$,$k$) using Eq. \ref{PathPairDifference}. The mean, standard deviation and variance of the path delay differences are then computed across all chips using Eqs. \ref{PathPairMean}, \ref{PathPairStdDev} and \ref{PathPairVariance}, respectively.

\begin{equation}
    \mathbf{PDCD_{i,(j,k)}} = (PDC_{i,j} - PDC_{i,k})
    \label{PathPairDifference}
\end{equation}

\begin{equation}
    \mathbf{u_{PDCD_{j,k}}} = \frac{\sum\limits_{i=1}^{NC}PDCD_{i,(j,k)}}{NC}
    \label{PathPairMean}
\end{equation}

\begin{equation}
    \mathbf{\sigma_{PDCD_{j,k}}} = \sqrt\frac{\sum\limits_{i=1}^{NC}(PDCD_{i,(j,k)} - \mathbf{u_{PDCD_{j,k}}})^2}{NC - 1}
    \label{PathPairStdDev}
\end{equation}

\begin{equation}
    \mathbf{\sigma^2_{PDCD_{j,k}}} = \mathbf{(\sigma_{PDCD_{j,k}})^2}
    \label{PathPairVariance}
\end{equation}

\subsection{Node Variation Analysis}\label{SubSection:NodeVariationAnalysis}

In this section, we leverage the \textit{L-LUT-Mismatch} classes and \textit{N-Node-Mismatch} subclasses of path pairings to derive an estimate of variance that is associated with the nodes in the FPGA's programmable logic. The mean values of the variances within each \textit{L-LUT-Mismatch} and \textit{N-Node-Mismatch} class are computed using Eq. \ref{MeanPathPairVariance}, with $r$ representing a member from the set of $(j,k)$ path pairings that meet the class $L$ and subclass $N$ conditions.

\begin{equation}
    \mathbf{u}_{\sigma^2_{L,N}} = \frac{\sum\limits_{r=1}^{NP_{L,N}}\mathbf{\sigma^2_{PDCD_{r}}}}{NP_{L,N}}
    \label{MeanPathPairVariance}
\end{equation}

The measured variance for path pairings from a \textit{L-LUT-Mismatch} and \textit{N-Node-Mismatch} (sub)class can be modeled as a combination of LUT and node variance as given by Eq. \ref{LUTNodeVarianceDifferenceSum}, where $L$ represents the number of distinct LUTs and $N$ represents the number of distinct nodes. Given this model, it is straightforward to derive the node variance by subtracting consecutive \textit{N-Node-Mismatch} subclasses as given by Eqs. \ref{NodeVariance1} and \ref{NodeVariance2}. We use this differencing technique substituting the mean measured values from Eq. \ref{MeanPathPairVariance} for the $\mathbf{\sigma^2_{L,N}}$ in Eq. \ref{LUTNodeVarianceDifferenceSum} and then solve for $\mathbf{u_{\sigma^2_N}}$.

\begin{equation}
    \mathbf{\sigma^2_{L,N}} = L*\sigma_L^2 + N*\sigma_{N}^2
    \label{LUTNodeVarianceDifferenceSum}
\end{equation}

\begin{equation}
    \mathbf{\sigma_{N}^2} = \sigma_{L,(N+1)}^2 - \sigma_{L,N}^2
    \label{NodeVariance1}
\end{equation}

\begin{equation}
    \mathbf{\sigma_{N}^2} = L*\sigma_{L}^2 + (N+1)*\sigma_{N}^2 - (L*\sigma_{L}^2 + N*\sigma_{N}^2)
    \label{NodeVariance2}
\end{equation}

The analysis is carried out on \textit{L-LUT-Mismatch} classes from 6, 8, 10 through 26. Within each of these classes, only those \textit{N-Node-Mismatch} subclasses with membership exceeding 200 and which did not posses any individual path pairs with variance classified as an outlier (beyond 3*$\sigma$ over that from other members of the class) are included. This down-selection process was necessary because of the large variance range that is associated with the individual paths as shown earlier in reference to Figs. \ref{Figure:SinglePathSortedVariance} and \ref{Figure:SinglePathSortedVarianceStats}. 

\begin{figure}[t]
    \centering
     \includegraphics[width=3.4in,keepaspectratio=true]{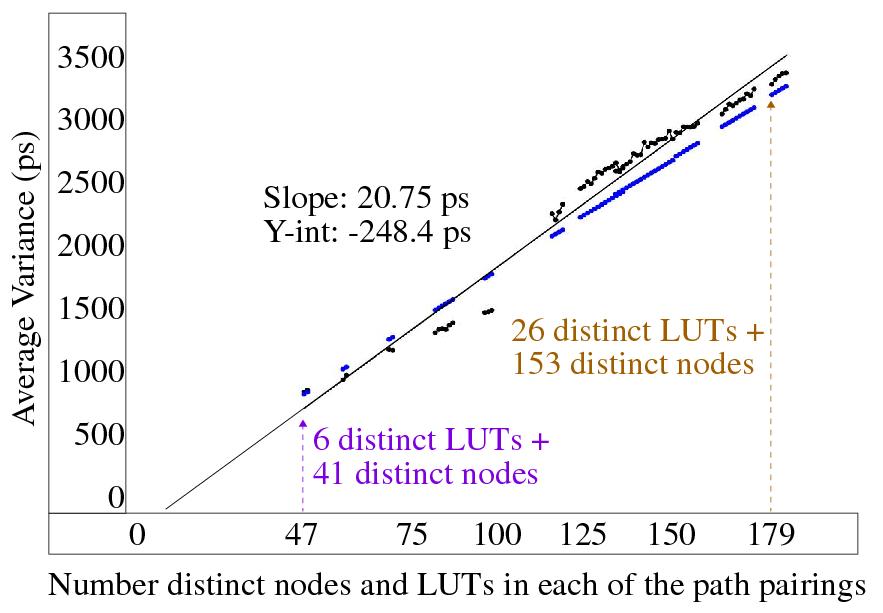}
     \caption{Node Analysis over LUT classes 6 through 26.}
     \label{Figure:NodeVarianceAnalysis}
     \vspace{-12pt}
\end{figure}

The results of this screening process are shown in Fig. \ref{Figure:NodeVarianceAnalysis}. Here, the black points represent the computed means $\mathbf{u_{\sigma^2_{L,N}}}$ plotted within each \textit{L-LUT-Mismatch} class as a set of consecutive points in which the number of nodes increases by one from left to right. Differences are computed between pairs of consecutive points within each \textit{L-LUT-Mismatch} class, in the spirit of Eq. \ref{NodeVariance2}, and then an average $\mathbf{u_{\sigma^2_N}}$ is computed for all point pair differences for that class. The final value for Node variance, computed as the mean of the means across all \textit{L-LUT-Mismatch} classes, is $\mathbf{u_{\sigma^2_N}} = 16.83$ ps. 

The black line is the best fit line (using linear regression) through the measured data (black points), and is included to illustrate that the set of measured values for $\mathbf{\sigma^2_{L,N}}$ as a group possess the desirable property of having a y-intercept close to 0.0. The slope and y-intercept are given as 20.75 ps and -248.4 ps.

A series of estimates of LUT variance can then be obtained using this estimate of the node variance by subtracting $N$ copies of the node variance from the measured values (black points) and then dividing by the number of LUTs in the class, as given by Eq. \ref{LUTvariance}. The final value for LUT variance is $\mathbf{u_{\sigma^2_L}} = 25.38$ ps, computed as the mean of the means across all \textit{L-LUT-Mismatch} classes.

\begin{equation}
    \mathbf{u_{\sigma_{L}^2}} = \mathbf{(u_{\sigma^2_{L,N}} - N * u_{\sigma^2_N})/L}
    \label{LUTvariance}
\end{equation}

The blue points in Fig. \ref{Figure:NodeVarianceAnalysis} represent the predicted values of $\mathbf{u_{\sigma^2_{L,N}}}$ for each of the measured black points using the above estimates of average node and LUT variance in Eq. \ref{NodeVariance2}. The estimates of $\mathbf{u_{\sigma^2_N}}$ and $\mathbf{u_{\sigma^2_L}}$ are also used to predict the single path variance presented earlier, and are shown as superimposed lines in Figs. \ref{Figure:SinglePathSortedVariance} and \ref{Figure:SinglePathGroupAvePredictionDataBoth}. The uncertainty associated with these estimates is bounded by the largest value in the 'variance of the variances' curve shown in Fig. \ref{Figure:SinglePathSortedVarianceStats}. The largest value is 6354 ps at x-axis value 154. This class includes an average of 24.3 LUTs and 129.7 nodes. The predicted overall variance under these conditions is 24.3*25.38 + 129.7*16.83 = 2800 ps. The ratio of the predicted variance to worst-case variance of the measured variances is 2.27. Therefore, the actual variance of a randomly chosen individual LUT or node can be more than $2x$ larger than the mean value. The large positive excursions of variance around the derived mean value reflects an increase in entropy, which is beneficial in a PUF application.

The variance estimates can be easily converted into ThreeSigma estimates, which better reflect Entropy as we pointed out earlier in Section \ref{SubSection:SinglePathVariationAnalysis}. In particular, the equivalent ThreeSigmas are given as follows: 

$\mathbf{u_{ThreeSigma_{LUT}}} = 15.1$ ps 

$\mathbf{u_{ThreeSigma_{Node}}} = 12.3$ ps

The mean length path has approximately 16 LUTs and 87 nodes, yielding an expected ThreeSigma variation of $3*\sqrt{16.5*25.38}$ = $\mathbf{\pm} 61.4$ ps for LUTs and $3*\sqrt{86.9*16.83}$ = $\mathbf{\pm} 114.7$ ps for nodes, and a total ThreeSigma of $3*\sqrt{16.5*25.38 + 86.9*16.83}$ = $\mathbf{\pm} 130.1$ ps.

\begin{figure}[t]
    \centering
     \includegraphics[width=3.4in,keepaspectratio=true]{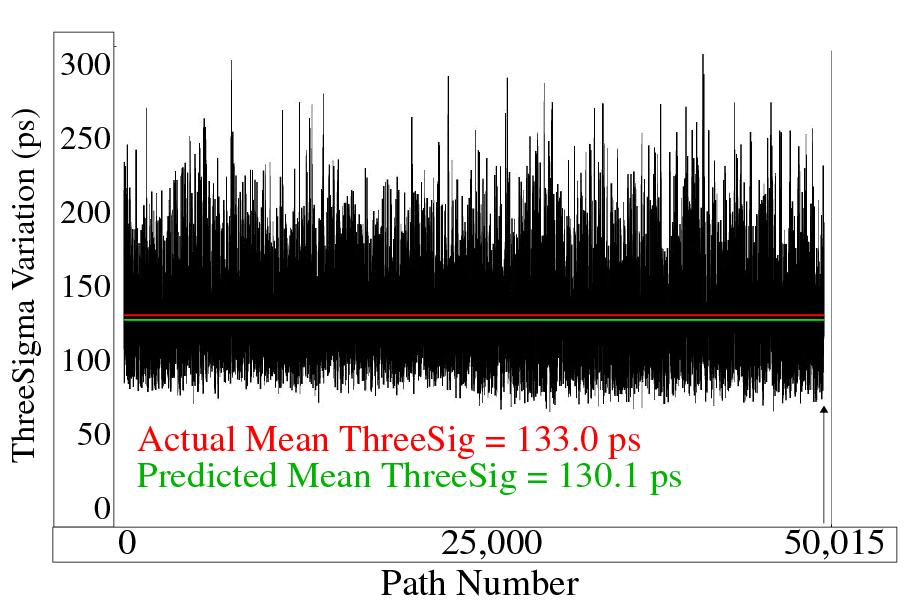}
     \caption{ThreeSigma variations of all 50,015 paths, with actual and predicted mean ThreeSigma variations superimposed as horizontal lines.}
     \label{Figure:SinglePathAllThreeSigmas}
     \vspace{-10pt}
\end{figure}

This predicted mean ThreeSigma is illustrated as a green horizontal line in Fig. \ref{Figure:SinglePathAllThreeSigmas}, superimposed on top of the Three sigmas for all 50,015 paths. The actual ThreeSigma is shown as a red horizontal line for comparison. The predicted mean ThreeSigma underestimates the actual variation by less than 3 picoseconds, which validates our estimation technique for node and LUT variance. As discussed above, individual ThreeSigma variations vary $\mp$ around this mean value. The smallest ThreeSigma is approx. 70 ps, slightly larger than \( \frac{1}{2} \)$x$ the mean value.

\subsection{Entropy to Bitstring Generation within HELP}\label{Section:NodeVariationAnalysis}

\begin{figure*}[t]
    \centering
     \includegraphics[width=4.0in,keepaspectratio=true]{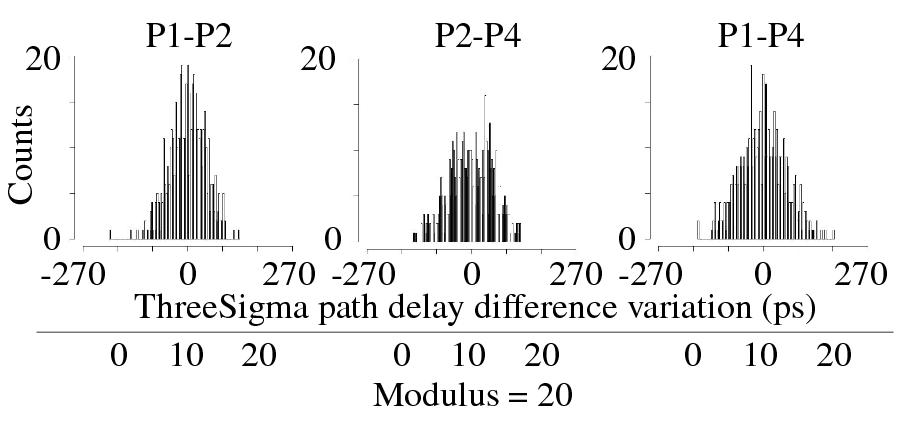}
     \caption{Plots of compensated path delay difference distributions from 500 PUF instances with mean subtracted ($PNCDO$) illustrating the level of entropy (in picoseconds) available for a PUF application. The HELP PUF applies a modulus operation to the fine phase shift representations of the compensated path delay differences, that maps the distributions to a range between 0 and 19.9, as shown along the bottom of the plots. Bitstring generation assigns a bit value of '0' for $PNCDO$ in the range 0 to 9.9 and a bit value of '1' for $PNCDO$ in the range of 10 to 19.9.}
     \label{Figure:P1_P2_P4_PathDIFFVariationDist}
     \vspace{-10pt}
\end{figure*}

In this section, we use the results of this analysis as support for the assessment reported in \cite{CheEntropy:2017} that HELP-generated bitstrings possess high statistical quality. The HELP bitstring generation algorithm post-processes the delay values using the same set of operations described in the previous sections, i.e., HELP computes path delay differences from individual path delay values and applies the linear transformations described in Section \ref{SubSection:VariationAndCompensation} to a distribution of size 2048 to create a set of $PDCD$. A modulus is applied to the $PDCD$ to remove path length bias, and offsets are applied as an optimization technique to improve the uniqueness statistical characteristic of the bitstrings. The offsets are computed by the server using data from a provisioned set of FPGAs, and are just the mean values of the $PDCD$ computed across all PUF instances. The offsets are stored as helper data, and are subtracted from the $PDCD$ to generate $PDCDO$ as given by Eq. \ref{PathPairDifferenceOptimized}. 

\begin{equation}
    \mathbf{PDCDO_{i,(j,k)}} = PDCD_{i,(j,k)} - u_{PDCD_{j,k}}
    \label{PathPairDifferenceOptimized}
\end{equation}

Fig. \ref{Figure:P1_P2_P4_PathDIFFVariationDist} shows a set of $PDCDO$ distributions using data from the 500 PUF instances. The individual path delays from Fig. \ref{Figure:P1_P2_P4_PathVariationDist} are used to create these three combinations of delay differences, namely P1-P2, P2-P4 and P1-P4. The distributions are created using Eq. \ref{PathPairDifferenceOptimized}, which computes delay differences for all PUF instances $i$ for path pair ($j$, $k$) using Eq. \ref{PathPairDifference} and then subtracts the mean, which is derived for each of the path pair differences using Eq. \ref{PathPairMean}.

Path P1 has 15 LUTs and 82 nodes, P2 has 19 LUTs and 107 nodes while P4 has 17 LUTs and 98 nodes, yielding ThreeSigma predictions given by the following expressions:

\begin{multline}
    \mathbf{ThreeSigma_{P1-P2}} = \\3*\sqrt{(15 + 19)*25.38 + (82 + 107)*16.83}
    \\= 190 ps
    \label{PathPairP1P2Example}
\end{multline}

\begin{multline}
    \mathbf{ThreeSigma_{P2-P4}} = \\3*\sqrt{(19 + 17)*25.38 + (107 + 98)*16.83}
    \\= 198 ps
    \label{PathPairP2P4Example}
\end{multline}

\begin{multline}
    \mathbf{ThreeSigma_{P1-P4}} = \\3*\sqrt{(15 + 17)*25.38 + (82 + 98)*16.83}
    \\= 186 ps
    \label{PathPairP1P4Example}
\end{multline}

As discussed earlier in Section \ref{Section:ExperimentalDesign}, the HELP algorithm measures, stores and processes path delays in units of fine phase shift (FPS), where 1 FPGA unit is equivalent to 18 ps. HELP applies a modulus operation to the path delay differences, which bounds the differences within the range of the modulus. For example, a modulus of 20 is used to re-label the x-axis in Fig. \ref{Figure:P1_P2_P4_PathDIFFVariationDist} in units of FPS. Path delay differences to the left of 10 in these figures are assigned a bit value of 0 while those shown to the right are assigned a bit value of 1. As discussed above, the HELP algorithm applies offsets, i.e., small shifts less than modulus/2, to the distributions to ensure that the difference distributions are always centered over modulus/2, as shown in the plots. These operations significantly improve both the randomness and uniqueness statistical properties of the generated bitstrings.

\section{Summary and Conclusions}\label{Section:Conclusions}
The basic components of an IC, namely, the gates and wires, provide the underlying source of Entropy for physical unclonable functions (PUFs). The electrical parameters of these components, e.g., wire resistance and delay, vary (within limits) because of the non-zero tolerance associated with the IC manufacturing process. This paper analyzes a large set of path delays collected from a set of Xilinx Zync 7020 FPGAs to derive an estimate of the average delay variation associated with the individual LUTs and nodes within the programmable fabric of the FPGAs. The application of linear compensation, statistical averaging and differencing techniques to the data collected from 500 identical copies of the HELP PUF's functional unit (source of entropy) enables accurate estimates of LUT and node variances to be obtained while minimizing artifacts related to measurement noise and global chip-to-chip variations in the measured delays. 

A set of vectors is applied to the FPGAs in hardware experiments to obtain the delays associated with a unique, but overlapping, set of 50,015 paths embedded within the PUF instances. A custom TCL script is developed and applied to the physical layout of the functional unit to extract the number and names of the LUTs and nodes that define each of the paths. This structural information is used to find pairing of paths in which some fraction of LUTs and nodes in the two paths of the pairing overlap. An algorithm is developed that partitions the complete set of 625 million path pairings into smaller subsets, each with a specific set of common LUT and node components. Statistical differencing and averaging over these sets enables the variance of individual LUTs and nodes to be estimated. The final estimates of individual LUT and node variance are obtained by averaging the results obtained from each of the set analyses. The estimates are then validated by comparing the measured variances of the individual paths with those predicted by our numerical technique. The analysis reveals that the range ($3*\sigma$) of delay variation for LUTs and nodes is approximately 15.1 ps and 12.3 ps, respectively.  

% use section* for acknowledgment
\section*{Acknowledgment}\label{Section:Acks}
This work was supported in part by Enthentica, Inc. and Raytheon, Inc.

\ifCLASSOPTIONcaptionsoff
  \newpage
\fi

\bibliographystyle{IEEEtran}
\bibliography{IEEEabrv,references.bib}

% Generated by IEEEtran.bst, version: 1.14 (2015/08/26)
\begin{thebibliography}{10}
\providecommand{\url}[1]{#1}
\csname url@samestyle\endcsname
\providecommand{\newblock}{\relax}
\providecommand{\bibinfo}[2]{#2}
\providecommand{\BIBentrySTDinterwordspacing}{\spaceskip=0pt\relax}
\providecommand{\BIBentryALTinterwordstretchfactor}{4}
\providecommand{\BIBentryALTinterwordspacing}{\spaceskip=\fontdimen2\font plus
\BIBentryALTinterwordstretchfactor\fontdimen3\font minus
  \fontdimen4\font\relax}
\providecommand{\BIBforeignlanguage}[2]{{%
\expandafter\ifx\csname l@#1\endcsname\relax
\typeout{** WARNING: IEEEtran.bst: No hyphenation pattern has been}%
\typeout{** loaded for the language `#1'. Using the pattern for}%
\typeout{** the default language instead.}%
\else
\language=\csname l@#1\endcsname
\fi
#2}}
\providecommand{\BIBdecl}{\relax}
\BIBdecl

\bibitem{Aarestad2013}
J.~Aarestad, P.~Ortiz, D.~Acharyya, and J.~Plusquellic, ``Help: A
  hardware-embedded delay puf,'' \emph{IEEE Design \& Test}, vol.~30, no.~2,
  pp. 17--25, 2013.

\bibitem{YuFPGA2010}
H.~Yu, Q.~Xu, and P.~H.~W. Leong, ``Fine-grained characterization of process
  variation in fpgas,'' in \emph{International Conference on Field-Programmable
  Technology}.\hskip 1em plus 0.5em minus 0.4em\relax IEEE, 2010, pp. 138--145.

\bibitem{WongReconfig2009}
\BIBentryALTinterwordspacing
J.~S.~J. Wong, P.~Sedcole, and P.~Y.~K. Cheung, ``Self-measurement of
  combinatorial circuit delays in fpgas,'' \emph{ACM Trans. Reconfigurable
  Technol. Syst.}, vol.~2, no.~2, Jun. 2009. [Online]. Available:
  \url{https://doi.org/10.1145/1534916.1534920}
\BIBentrySTDinterwordspacing

\bibitem{mehta2012limit}
N.~Mehta, R.~Rubin, and A.~DeHon, ``Limit study of energy \& delay benefits of
  component-specific routing,'' in \emph{Proceedings of the ACM/SIGDA
  international symposium on Field Programmable Gate Arrays}, 2012, pp.
  97--106.

\bibitem{gojman2014grokint}
B.~Gojman and A.~DeHon, ``Grok-int: Generating real on-chip knowledge for
  interconnect delays using timing extraction,'' in \emph{2014 IEEE 22nd Annual
  International Symposium on Field-Programmable Custom Computing
  Machines}.\hskip 1em plus 0.5em minus 0.4em\relax IEEE, 2014, pp. 88--95.

\bibitem{gojman2014groklab}
B.~Gojman, S.~Nalmela, N.~Mehta, N.~Howarth, and A.~DeHon, ``Grok-lab:
  Generating real on-chip knowledge for intra-cluster delays using timing
  extraction,'' \emph{ACM Transactions on Reconfigurable Technology and Systems
  (TRETS)}, vol.~7, no.~4, pp. 1--23, 2014.

\bibitem{Schaumont2010}
A.~Maiti, J.~Casarona, L.~McHale, and P.~Schaumont, ``A large scale
  characterization of ro-puf,'' in \emph{2010 IEEE International Symposium on
  Hardware-Oriented Security and Trust (HOST)}, 2010, pp. 94--99.

\bibitem{Kaps:2013}
B.~Habib, K.~Gaj, and J.~Kaps, ``Fpga puf based on programmable lut delays,''
  \emph{Euromicro Conference on Digital System Design}, 2013.

\bibitem{CheEntropy:2017}
W.~Che, V.~K. Kajuluri, M.~Martin, F.~Saqib, and J.~Plusquellic, ``Analysis of
  entropy in a hardware-embedded delay puf,'' \emph{Cryptography}, vol.~1,
  no.~1, 2017.

\bibitem{Tiri:2004}
K.~Tiri and I.~Verbauwhede, ``A logic level design methodology for a secure dpa
  resistant asic or fpga implementation,'' in \emph{Proceedings Design,
  Automation and Test in Europe Conference and Exhibition}, 2004, pp. 246--251.

\end{thebibliography}

\begin{IEEEbiography}[{\includegraphics[width=1in,height=1.25in,clip,keepaspectratio]{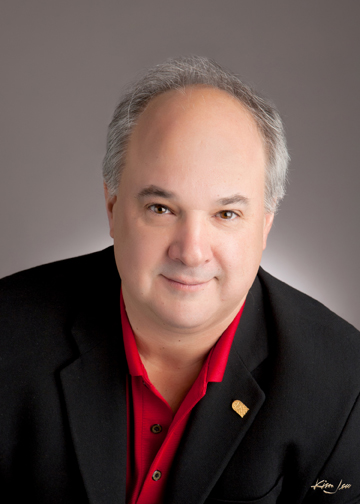}}]{Jim Plusquellic}
is a Professor in Electrical and Computer Engineering at the University of New Mexico. He received both his M.S. and Ph.D. degrees in Computer Science from the University of Pittsburgh. Professor Plusquellic received an "Outstanding Contribution Award" from IEEE Computer Society in 2012 and 2017 for co-founding and for his contributions to the Symposium on Hardware-Oriented Security and Trust (HOST). He is the Trust and Assurance Theme Lead for ASU's SWAP Hub.
\end{IEEEbiography}

\begin{IEEEbiography}[{\includegraphics[width=1in,height=1.25in,clip,keepaspectratio]{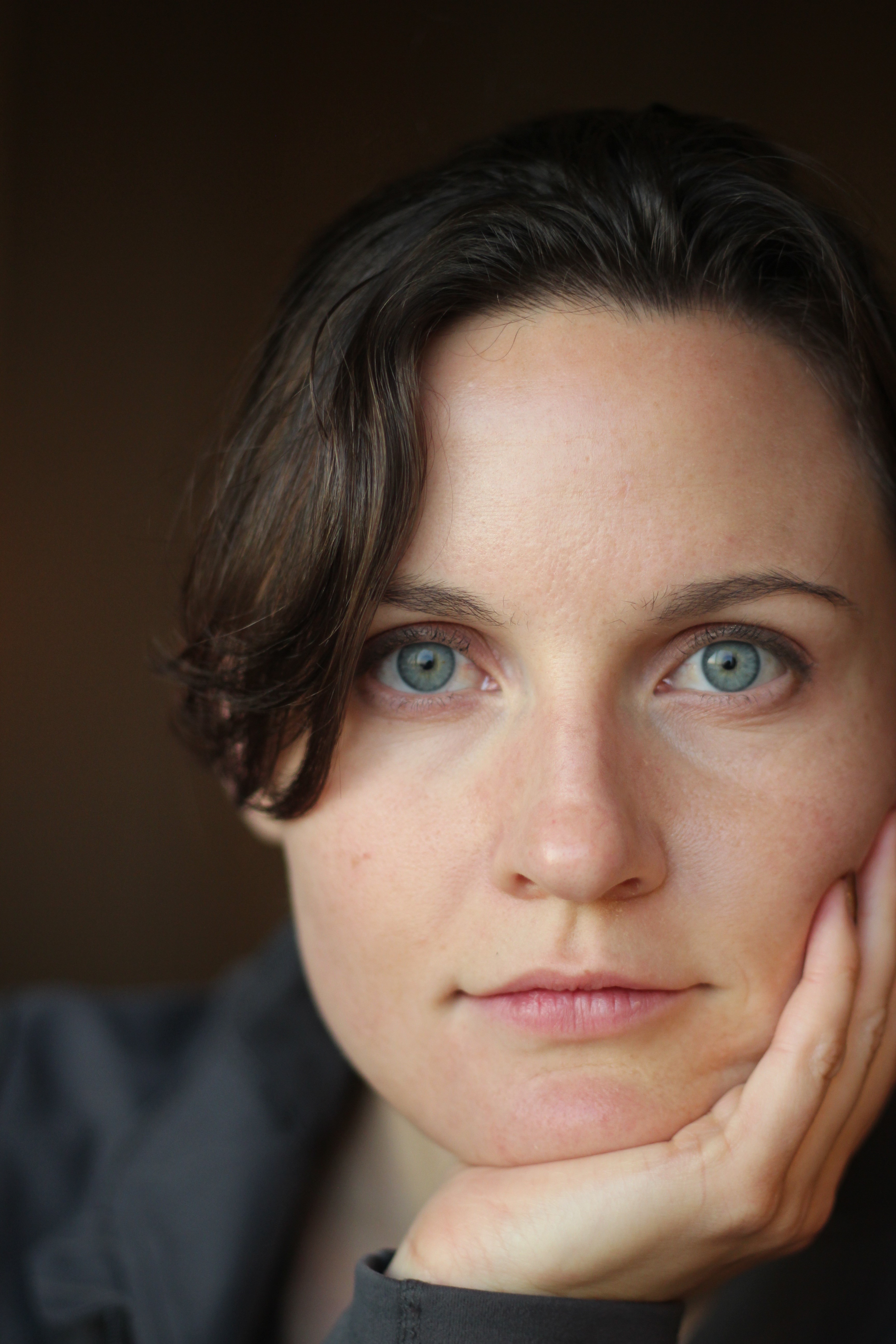}}]{Jennifer Howard} 
is a Systems Engineer at Martin UAV (a Shield AI company). She received both her B.S. and M.S. in Electrical Engineering from the University of Idaho. 
\end{IEEEbiography}

\begin{IEEEbiography}[{\includegraphics[width=1in,height=1.25in,clip,keepaspectratio]{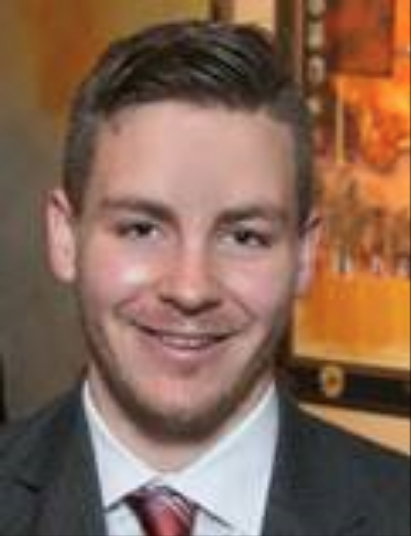}}]{Ross MacKinnon}
Ross A. MacKinnon is a Systems Engineer at Martin UAV (a Shield AI Company) with a background in embedded real-time systems and system security. He received his B.S. in Electrical Engineering from the University of Southern California and his M.S. in Systems Engineering from Johns Hopkins University.
\end{IEEEbiography}

\begin{IEEEbiography}[{\includegraphics[width=1in,height=1.25in,clip,keepaspectratio]{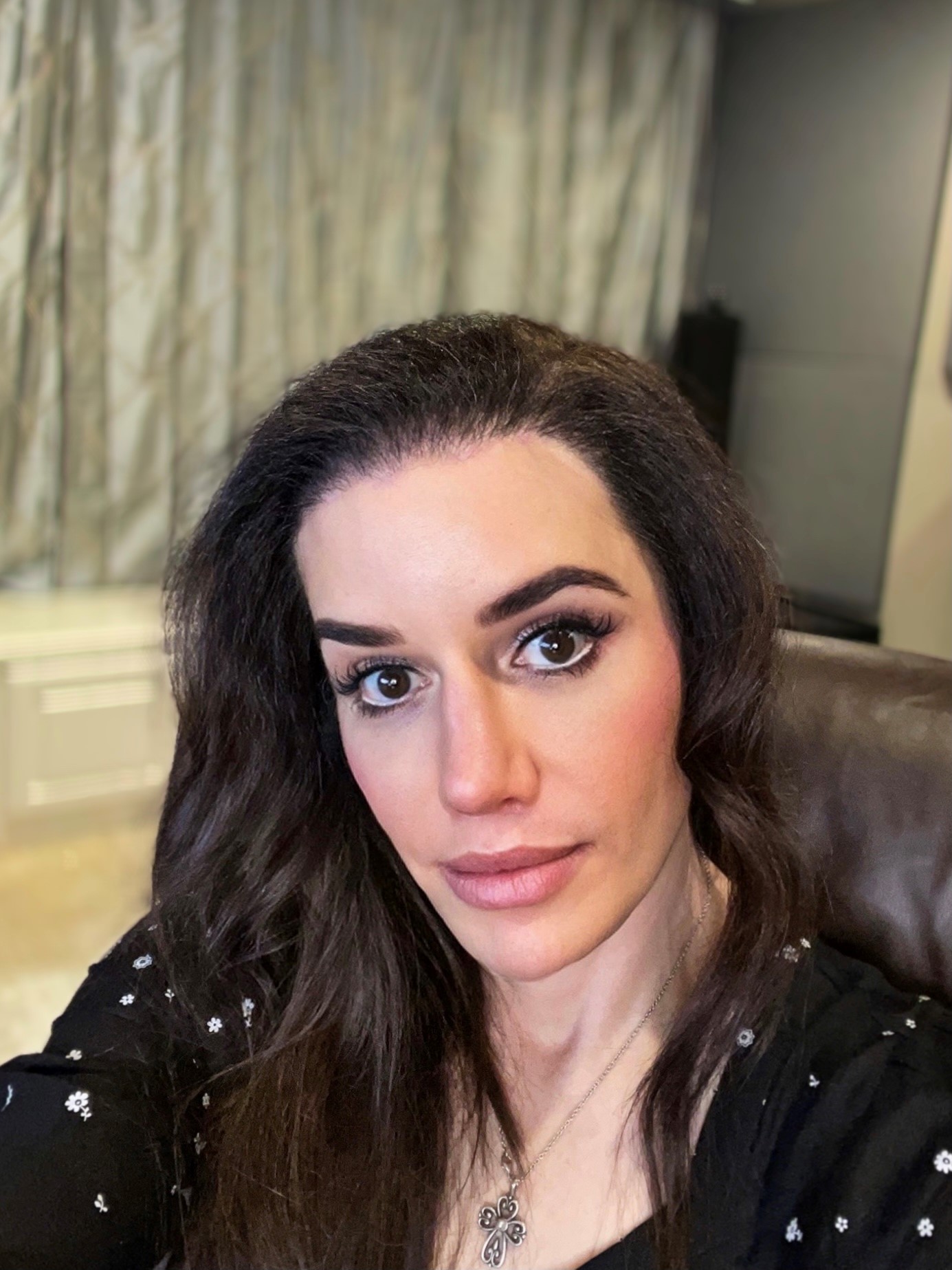}}]{Kristianna Hoffman}
is a Principle Engineering Fellow and Certified Raytheon and Open Group Architect with over twenty years of experience.  Kristi holds a Bachelor's in Electrical Engineering from Texas Tech University, and a Master's and Ph.D. in Computer Architecture and VLSI from the University of New Mexico. 
\end{IEEEbiography}

\begin{IEEEbiography}[{\includegraphics[width=1in,height=1.25in,clip,keepaspectratio]{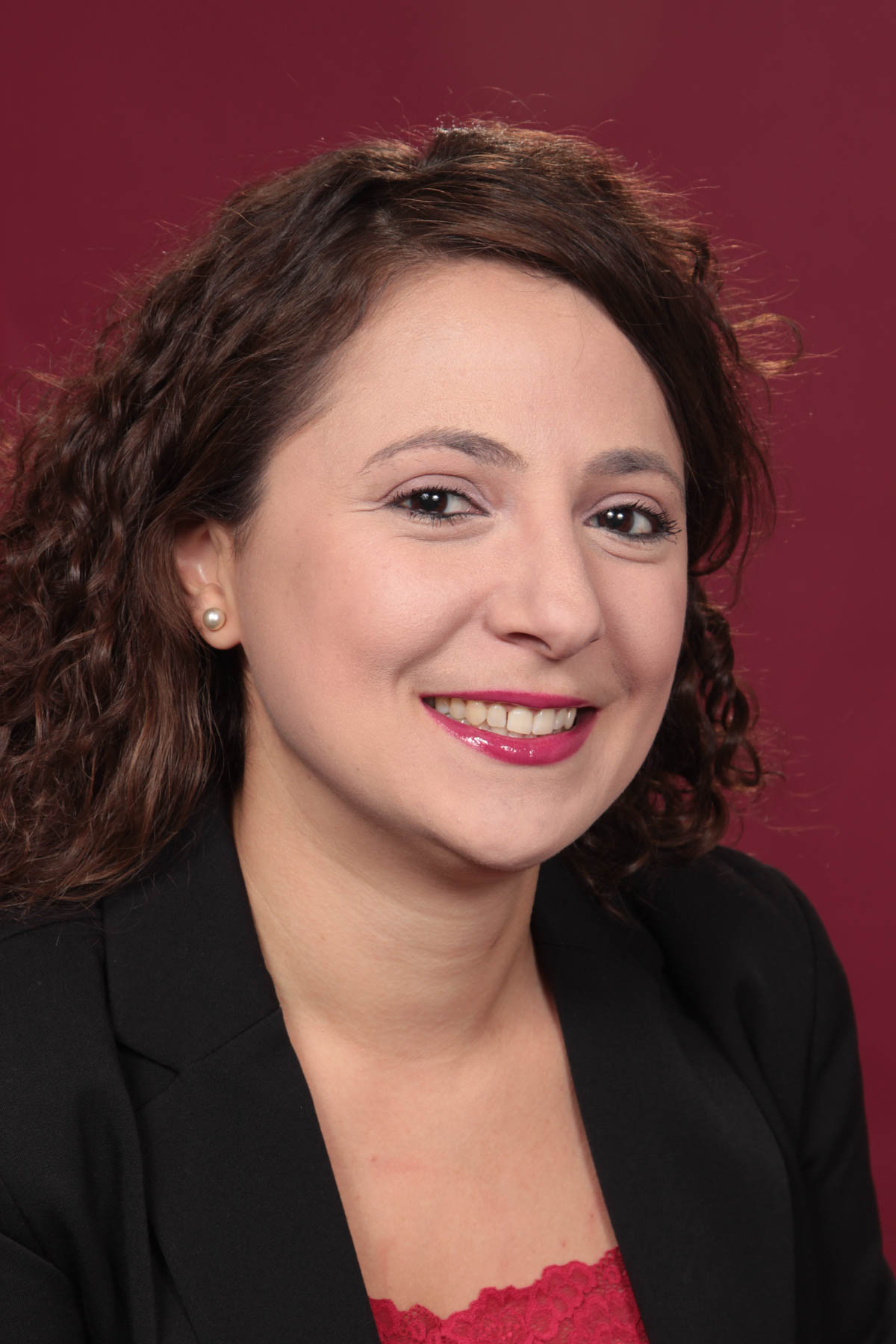}}]{Eirini Eleni Tsiropoulou} is currently an Assistant Professor at the Department of Electrical and Computer Engineering, University of New Mexico. Her main research interests lie in the area of cyber-physical social systems and wireless heterogeneous networks, with emphasis on network modeling and optimization, resource orchestration in interdependent systems, reinforcement learning, game theory, network economics, and Internet of Things. Four of her papers received the Best Paper Award at IEEE WCNC in 2012, ADHOCNETS in 2015, IEEE/IFIP WMNC 2019, and INFOCOM 2019 by the IEEE ComSoc Technical Committee on Communications Systems Integration and Modeling. She was selected by the IEEE Communication Society - N2Women - as one of the top ten Rising Stars of 2017 in the communications and networking field. She received the NSF CRII Award in 2019 and the Early Career Award by the IEEE Communications Society Internet Technical Committee in 2019.
\end{IEEEbiography}

\begin{IEEEbiography}[{\includegraphics[width=1in,height=1.25in,clip,keepaspectratio]{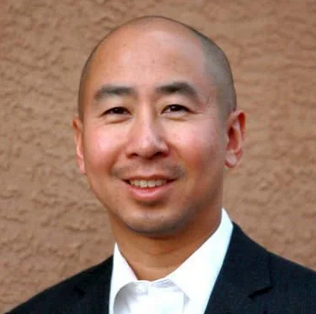}}]{Calvin Chan} Senior Research Associate, Center for National Security Initiatives, University of Colorado Boulder, Boulder CO 
\end{IEEEbiography}

\end{document}